\let\TeXprimitiveyear\year
\documentclass{ieeeaccess}
\makeatletter

\let\year\TeXprimitiveyear
\providecommand{\xfigwd}{\columnwidth}
\makeatother

\definecolor{accessblue}{rgb}{0,0.384,0.608}
\providecommand{\xfigwd}{\columnwidth}
\makeatother
\usepackage{cite}
\usepackage{amsmath,amssymb,amsfonts}
\usepackage{algorithmic}
\usepackage{graphicx}
\usepackage{textcomp}
\usepackage{xspace}
\usepackage{url}

\usepackage{framed}
\usepackage{subfigure}

\usepackage{siunitx}
\usepackage{adjustbox}
\usepackage{array}
\usepackage{multirow}
\usepackage{booktabs}
\usepackage{colortbl}
\usepackage{lscape}
\usepackage{url}
\usepackage{xspace}
\usepackage{framed}
\usepackage{subfigure}
\usepackage{tcolorbox}
\usepackage{pgf}
\usepackage{xcolor}
\tcbuselibrary{listingsutf8}
\usepackage{upquote}

\def\BibTeX{{\rm B\kern-.05em{\sc i\kern-.025em b}\kern-.08em
    T\kern-.1667em\lower.7ex\hbox{E}\kern-.125emX}}

\def\tsc#1{\csdef{#1}{\textsc{\lowercase{#1}}\xspace}}
\tsc{WGM}
\tsc{QE}
\tsc{EP}
\tsc{PMS}
\tsc{BEC}
\tsc{DE}

\newcommand{\newcolor}{black}
\newcommand{\new}[1]{#1}
\newcommand{\newt}[1]{\textcolor{\newcolor}{#1}\xspace}

\newcommand{\qoslanguage}{Quantum QoS Specification Language\xspace}
\newcommand{\qoslanguageacr}{QQSpec\xspace}

\definecolor{myblue}{rgb}{0.9,0.95,1} 
\newcounter{myboxcounter}[section]

\newtcolorbox[use counter=myboxcounter, number within=section]{ebnfbox}[2][]{%
  colback=myblue,
  colframe=cyan!70!black,
  fonttitle=\bfseries,
  coltitle=black,
  listing only,
  title={Listing~\themyboxcounter:~#2},
  left=3mm,
  right=3mm,
  top=2mm,
  bottom=2mm,
  boxrule=0.6mm,
  arc=2mm,
  float,
  #1,
  label={#2},
}

\newtcolorbox[use counter=myboxcounter, number within=section]{modelview}[2][]{%
  colback=cyan!6!white,
  colframe=cyan!70!black,
  fonttitle=\bfseries,
  coltitle=black,
  title={Listing~\themyboxcounter:~#2},
  arc=2mm,
  boxrule=1pt,
  #1
}

\usepackage{xcolor}

\newcommand{\rev}[1]{\textcolor{black}{#1}}

\newenvironment{revision}
{\color{black}}
{}

\begin{document}

\history{Date of publication xxxx 00, 0000, date of current version xxxx 00, 0000.}
\doi{10.1109/TQE.20XX.XXXXXXX}

\title{Quantum Broker: QoS-Aware Distribution of Quantum Computations}

\author{\uppercase{Giuseppe Bisicchia}\authorrefmark{1,2},
\uppercase{Alessandro Bocci}\authorrefmark{1},
\uppercase{Jose Garcia-Alonso}\authorrefmark{2},
\uppercase{Juan Manuel Murillo}\authorrefmark{2}, and
\uppercase{Antonio Brogi}\authorrefmark{1}}

\address[1]{University of Pisa, Pisa, Italy (e-mail: giuseppe.bisicchia@phd.unipi.it; alessandro.bocci@unipi.it; antonio.brogi@unipi.it)}
\address[2]{University of Extremadura, C\'{a}ceres, Spain (e-mail: jgaralo@unex.es; juanmamu@unex.es)}


\markboth
{Bisicchia \headeretal: Quantum Broker: QoS-Aware Distribution of Quantum Computations}
{Bisicchia \headeretal: Quantum Broker: QoS-Aware Distribution of Quantum Computations}

\corresp{Corresponding author: Giuseppe Bisicchia (e-mail: giuseppe.bisicchia@phd.unipi.it).}

\begin{abstract}
Cloud-accessible quantum computing has made heterogeneous Quantum Processing Units (QPUs) available through provider platforms that differ in qubit technology, calibration state, queueing behavior, pricing model, access policy, and exposed metadata. This heterogeneity makes execution management a quantum-engineering concern: a quantum circuit may be executable on several backends, but the resulting cost, latency, fidelity, and reproducibility depend on backend selection and on how measurement shots are allocated. This paper presents the \textit{Quantum Broker}, an open-source and provider-agnostic framework for Quality-of-Service-aware quantum execution management. The framework operationalizes shot-wise distribution by exploiting the statistical independence of quantum measurement shots to dispatch a single circuit across multiple heterogeneous QPUs according to user-defined constraints and optimization objectives. It combines a modular execution architecture, interchangeable backend-profiling strategies, and the \textit{Quantum QoS Specification Language} (QQSpec), a declarative language for expressing requirements on cost, execution time, waiting time, fidelity, backend usage, and shot allocation. \rev{Dispatch planning is supported by two solver backends: a
mixed-integer linear-programming (MILP) solver for efficient planning
under linearised QoS assumptions, and a nonlinear dual-annealing solver
for profile-driven optimisation under more complex QoS models.} Experiments on benchmark quantum circuits and noisy simulator backends show that the Quantum Broker can produce feasible QoS-aware dispatch plans under heterogeneous backend conditions and can improve QoS satisfaction with respect to single-backend and naive multi-backend baselines. The results also expose a practical trade-off between planning time and dispatch quality, with the nonlinear solver producing higher-quality allocations at increased computational cost. Overall, the Quantum Broker provides an extensible execution-management layer for engineering portable, reproducible, and QoS-aware quantum software over fragmented quantum cloud infrastructures.
\end{abstract}


\titlepgskip=-15pt

\maketitle


\section{Introduction}
\label{sec:intro}

\new{Quantum computing has moved from isolated laboratory prototypes to cloud-accessible quantum computing systems in which users submit circuits to Quantum Processing Units (QPUs) with different physical implementations, control stacks, calibration regimes, and service policies~\cite{Kim2023,Arute2019,quantumhandcrafting,kaiiali2019cloud}. Public quantum platforms now expose devices based on superconducting circuits, trapped ions, neutral atoms, and other qubit technologies, through providers such as IBM, IonQ, Rigetti, and OQC~\cite{serrano2022quantumsoftware}. This transition creates a quantum-engineering challenge at the boundary between quantum software, provider platforms, and heterogeneous Noisy Intermediate-Scale Quantum (NISQ) devices: the same circuit can be executable on multiple backends, but with different execution latencies, queueing delays, costs, error profiles, access constraints, and output quality~\cite{proctor2025benchmarking,lubinski2023application}.}

\new{For cloud-based quantum applications, execution management is therefore not a secondary deployment concern. It is part of the engineering of the quantum computing stack. Developers must often decide where and how to execute a circuit while accounting for limited qubit counts, short coherence times, restricted connectivity, gate and readout errors, finite-shot statistical uncertainty, and rapidly changing backend availability~\cite{preskill2018quantum}. In practice, these decisions are still frequently handled through provider-specific workflows: developers inspect backend metadata manually, select a device, adapt code to the selected provider API, and accept that the selected backend may become unsuitable when calibration data, queue state, pricing, or access policies change.}

\new{This situation exposes a gap between user-level execution intent and the operational behavior of quantum cloud infrastructures. Quantum software engineers may need to minimize execution time, bound monetary cost, constrain expected fidelity, avoid specific providers or regions, distribute load across several QPUs, or obtain reproducible execution plans for later comparison. However, current workflows provide limited support for expressing such Quality-of-Service (QoS) requirements in a portable form and for enforcing them automatically across heterogeneous providers. As a result, non-functional requirements that should be explicit engineering constraints remain embedded in ad hoc scripts, manual backend choices, or provider-specific configuration parameters.}

\new{The problem is amplified by the fragmentation of the current quantum cloud ecosystem. Providers expose different APIs, metadata schemas, pricing models, queueing mechanisms, and device-characterization reports. Moreover, quantum execution is intrinsically statistical: a job is typically evaluated through repeated shots, and the number and placement of those shots affect execution time, cost, and the quality of the empirical output distribution. These characteristics make QoS-aware execution a first-class quantum-engineering problem, requiring abstractions that connect requirement specification, backend profiling, optimization, and provider-agnostic orchestration.}

\new{Accordingly, this work addresses the following research question:}

\begin{quote}
\textit{How can quantum software engineers and developers specify and enforce Quality-of-Service requirements for quantum executions across heterogeneous, multi-provider quantum infrastructures in a portable, adaptive, and reproducible way?}
\end{quote}

\new{We focus on the execution phase of quantum software, where backend selection, shot allocation, workload distribution, and runtime adaptation determine whether a quantum task can satisfy practical engineering constraints. Rather than treating execution as the submission of a fixed circuit to a single backend, we model it as a QoS-aware dispatch problem. In this model, user-defined constraints, optimization objectives, backend profiles, and execution plans are represented explicitly and reasoned upon automatically.}

\new{In previous papers~\cite{shotwiseicsoc,shotwisenoisyarxiv}, we introduced the \textit{shot-wise distribution} methodology, which exploits the statistical independence of quantum measurement shots to distribute the execution of a single quantum circuit across multiple heterogeneous QPUs. \cite{shotwiseicsoc} presented also an initial \textit{Proof-of-Concept} (PoC) for automating shot distribution according to user-defined policies and backend characteristics.}

\new{In this manuscript, we evolve that proof of concept into the \textit{Quantum Broker}, an open-source software framework for QoS-aware quantum execution management. The framework treats quantum execution as an engineering process involving requirement specification, architectural abstraction, backend profiling, constraint-based reasoning, and provider-agnostic orchestration. Through a quantum-tailored \textit{\qoslanguage} (\qoslanguageacr), the Quantum Broker aims to close the gap between user intent and practical execution on fragmented, heterogeneous quantum cloud infrastructures.}

\rev{The novelty boundary with respect to our previous work is summarized in Table~\ref{tab:novelty}. In particular, this manuscript does \emph{not} claim shot-wise distribution itself as a new contribution: the underlying split/merge methodology, calibration concepts, incremental execution mechanisms, and an initial strategy-oriented DSL were developed in~\cite{shotwiseicsoc,shotwisenoisyarxiv}. The new contribution is the engineering of those ideas into an integrated QoS-aware execution-management framework in which user requirements, backend profiles, optimization-based reasoning, and provider-agnostic execution are connected through explicit software abstractions.}

\begin{table}[t]
\caption{\rev{Novelty of the present manuscript relative to our previous shot-wise work.}}
\label{tab:novelty}
\centering
\footnotesize
\setlength{\tabcolsep}{3pt}
\renewcommand{\arraystretch}{1.08}

\begin{revision}
\begin{tabular}{
    p{0.17\columnwidth}
    p{0.34\columnwidth}
    p{0.38\columnwidth}
}
\toprule
\textbf{Dimension} &
\textbf{Previous work~\cite{shotwiseicsoc,shotwisenoisyarxiv}} &
\textbf{This manuscript} \\
\midrule

Core scope &
Initial PoC (2023); split/merge methodology, calibration, and incremental execution (2025). &
Integrated Quantum Broker for QoS-aware execution management; shot-wise distribution is reused rather than claimed as new. \\

Requirements &
Strategy-oriented DSL and user-defined distribution policies. &
QQSpec: explicit QoS constraints/objectives over cost, time, waiting, quality/reliability, backend use, and shots. \\

Architecture &
Prototype/methodological workflow centered on distribution policies. &
Provider-agnostic Profiler--Reasoner--Executor broker architecture. \\

Decision making &
Policy/weight-based allocation and adaptive updates. &
Explicit optimization-based Reasoner: MILP for linearized QoS and nonlinear dual annealing for complex profiles. \\

Evaluation focus &
Split/merge behaviour, calibration, robustness, and output-distribution quality under noise. &
QoS feasibility/quality, single- and multi-backend baselines, and planning-time versus dispatch-quality trade-offs, complemented by end-to-end validation of Reasoner-selected dispatches through backend-specific execution, count aggregation, and comparison of the final application-visible distribution with an ideal reference ($F_{\mathrm{E2E}}$). \\

\bottomrule
\end{tabular}
\end{revision}

\end{table}

\rev{Against this prior-work baseline, the present manuscript makes the following contributions:}

\begin{itemize}
    \item \rev{\textbf{A dedicated QoS specification model.}
    We introduce \qoslanguageacr, a domain-specific language for expressing quantum execution constraints and optimization objectives, including cost, execution time, waiting time, fidelity, backend usage, and shot allocation.}

    \item \new{\textbf{A modular software architecture for quantum execution management.}
    We define a provider-agnostic architecture that separates requirement specification, profiling, reasoning, and execution concerns, enabling extensibility across providers, backends, solvers, and profiling strategies.}

    \item \new{\textbf{QoS-aware shot-wise dispatch.}
    We operationalize shot-wise distribution as a constrained optimization problem, allowing the shots of a single quantum circuit to be allocated across multiple heterogeneous QPUs according to user-defined QoS requirements.}

    \item \new{\textbf{Pluggable reasoning and profiling mechanisms.}
    We implement and compare two solver backends: a \rev{mixed-integer linear-programming (MILP) solver for fast dispatch planning under linearized QoS surrogates, with integer shot-allocation and binary backend-use variables}, and a nonlinear dual-annealing solver for profile-driven optimization under more complex QoS models. The profiling layer supports interchangeable strategies, including static profiles, provider metadata, and empirical benchmarking.}

    \item \new{\textbf{Empirical evaluation under controlled backend heterogeneity.}
    We evaluate the framework on benchmark quantum circuits and noisy simulator backends with fixed QoS profiles, comparing QoS-aware dispatch against exhaustive single-backend baselines and a naive multi-backend baseline.}

    \item \rev{\textbf{End-to-end validation of distributed quantum execution.}
    We complement the planning-level evaluation with an end-to-end experiment in which the total shot budget of a quantum circuit is distributed across multiple device-specific noisy simulator backends according to the selected execution plan. Each backend executes its assigned subset of shots, and the resulting measurement counts are then combined to reconstruct the final output distribution returned to the application. We compare this aggregate distribution with an independently generated ideal noise-free reference. This allows us to evaluate the quality of the final application-visible result after distributed execution and aggregation, rather than relying exclusively on the QoS estimates used to select the execution plan.}

    \item \new{\textbf{Open-source implementation.}
    We release the Quantum Broker as an open-source software package\footnote{\url{https://github.com/di-unipi-socc/TheQuantumBroker}}, enabling transparent reuse, community contribution, and reproducibility.}
\end{itemize}

\new{By framing quantum execution management as a QoS-aware quantum-engineering problem, this work contributes to the engineering of quantum software stacks and user platforms for cloud-accessible QPUs. The proposed framework helps developers express non-functional requirements independently of provider-specific APIs, reason explicitly about execution trade-offs, and obtain reproducible dispatch plans over heterogeneous quantum resources. In doing so, it supports a more systematic engineering practice for developing, executing, and evaluating quantum software in cloud-based environments.}

\newt{In this work, we distinguish between the design of a QoS-aware execution-management framework and the production benchmarking of live quantum-cloud services. Our primary objective is to study how QoS requirements can be specified, represented, and translated into dispatch plans over heterogeneous backend profiles. Accordingly, the empirical evaluation is conducted under controlled, real-device-inspired backend heterogeneity using noisy simulator backends and fixed QoS profiles. This choice enables reproducible comparison of dispatch strategies and solver behavior, while real-hardware deployment and calibration of profile parameters are treated as supported deployment paths and future validation steps.}

\rev{To additionally validate the behavior of the framework beyond dispatch planning, we perform an end-to-end execution experiment on the same class of device-specific noisy simulator backends. For the selected dispatches, each backend executes the exact number of shots assigned by the Reasoner; the returned counts are then merged to reconstruct the final shot-wise output distribution. We compare this aggregate distribution with an ideal noise-free reference, thereby evaluating the execution and aggregation stages independently from the synthetic QoS estimates used during optimization. This complementary experiment does not replace live-QPU validation, but it allows us to exercise and assess the complete dispatch--execution--aggregation workflow under reproducible backend-specific noise conditions.}

\new{The remainder of this paper is structured as follows. Section~\ref{sec:background} provides foundational background on quantum computing and introduces the main concepts of the \textit{shot-wise distribution} methodology and the optimization methods used in this work. Section~\ref{sec:related} surveys related work in quantum execution management and quantum QoS-aware scheduling. Section~\ref{sec:broker} details the design and implementation of the extended Quantum Broker. Section~\ref{sec:experiments} presents the experimental evaluation and discusses its main results. Section~\ref{sec:limitations} discusses current limitations, while Section~\ref{sec:concl} concludes the manuscript and outlines directions for future work.}

\section{Background}
\label{sec:background}

\subsection{Quantum Computing Execution}


A quantum computation is typically described by a \emph{quantum circuit}, which is a sequence of quantum gates (operations) applied to one or more qubits, followed by one or more measurements. These gates manipulate the qubit states, gradually evolving them into a desired superposition or entangled configuration. At the end of the circuit, measurements collapse the quantum state, yielding a classical output: a binary string corresponding to the observed state of the qubits.

Importantly, quantum circuits do not produce deterministic outputs. Instead, each execution of a circuit produces one random sample from a probability distribution over possible outputs. The probability of each outcome is determined by the final quantum state of the system before measurement. To recover this distribution and obtain statistically meaningful results, the same circuit must be executed many times. Each independent execution is called a \emph{shot}, and the collection of outcomes from multiple shots forms the empirical output distribution of the circuit.

Due to this probabilistic nature, quantum software must be designed with statistical post-processing in mind. Most quantum algorithms rely on analyzing the frequencies of observed outcomes across many shots to estimate properties such as expectation values, solution probabilities, or distributions over search spaces.

Complicating matters further is the presence of \emph{noise}. Current quantum hardware, often referred to as Noisy Intermediate-Scale Quantum (NISQ) devices, is susceptible to a range of physical imperfections (such as decoherence, gate errors, and readout inaccuracies) that distort the ideal quantum evolution and measurement outcomes. This noise introduces systematic alterations in the output distribution, reducing the fidelity of computation and making it harder to distinguish correct results from spurious ones. For a more in-depth discussion, we refer readers to standard resources such as the book by Nielsen and Chuang~\cite{nielsen2010quantum}.

\subsection{Shot-wise Distribution}
\textit{Shot-wise distribution} is a recently introduced quantum execution methodology that enables the parallel execution of a single quantum circuit across multiple quantum devices by distributing individual measurement shots among them~\cite{shotwiseicsoc,shotwiseqce}. \rev{The approach capitalises on the fact that repeated shots are statistically independent executions of the circuit, which permits the shot budget to be partitioned and scheduled independently across QPUs. Independence, however, is not by itself sufficient to guarantee an unbiased estimate of the ideal circuit output when heterogeneous QPUs have different effective noise channels. If every participating QPU samples from the same effective output distribution $p$ (for example, in the ideal/noiseless case or after an assumed correction that makes the device distributions statistically equivalent), pooling independent counts remains an unbiased estimator of that common $p$. In the general noisy case, if QPU $m$ produces distribution $p_m$ and receives $n_m$ of $N=\sum_m n_m$ shots, simple count pooling estimates the mixture $p_{\mathrm{pool}}=\sum_m (n_m/N)p_m$, which need not equal the ideal distribution. Thus, shot independence justifies splitting the sampling work, whereas bias with respect to the ideal target depends on the device-specific distributions and on the merge policy~\cite{shotwisenoisyarxiv}.}

This paradigm is particularly well-suited to the constraints of the NISQ era~\cite{preskill2018quantum}, in which devices are limited in qubit count, gate fidelity, and coherence time, and access to hardware is subject to queueing delays and variable availability. In such environments, submitting all measurement shots to a single QPU often results in high latency, exposure to device-specific noise profiles, and execution failure. Shot-wise distribution mitigates these challenges by enabling the concurrent execution of disjoint shot batches across multiple devices. If a device fails, only its assigned shots are affected, preserving the integrity of the remaining computation. Additionally, this strategy improves overall throughput and responsiveness, especially in cloud-based infrastructures where heterogeneous QPUs coexist and queue times fluctuate dynamically~\cite{shotwiseicsoc}.

The general execution model of shot-wise distribution comprises four main phases: (1) selecting a pool of candidate QPUs based on static attributes (such as qubit number, topology, and technology) and dynamic metrics (such as expected queue time or backend availability), (2) allocating the total shot budget across these devices according to a chosen policy (this can be uniform, weighted, or adaptive to real-time system conditions), (3) executing the assigned shots on each QPU in parallel, and (4) aggregating the returned measurement results into a unified output distribution. \rev{The split policy and the merge policy are conceptually distinct. Our default interpretation of a simple merge is raw count pooling, equivalent to weighting each QPU distribution by its executed-shot fraction $n_m/N$. This policy is appropriate when the participating QPUs can reasonably be treated as sampling a common effective distribution; otherwise it returns a shot-weighted mixture of their noisy distributions. When QPU reliabilities differ materially, the merge can instead use calibrated reliability-aware convex weights, statistical correction, or other post-processing. In particular, the shot-wise framework of~\cite{shotwisenoisyarxiv} explicitly studies uniform, Hellinger-based, and MISE-based merge strategies, with the latter policies using reliability information and, for MISE, balancing bias and variance. The Quantum Broker therefore treats aggregation as a configurable policy rather than assuming that independence alone makes heterogeneous outputs interchangeable.}

From a qualitative standpoint, the shot-wise methodology introduces significant improvements in robustness, flexibility, and developer control. Its decentralised execution model reduces the likelihood that a single backend bottleneck will compromise the entire computation, while its policy-driven architecture allows seamless adaptation to a wide range of user goals. Quantitatively, the experiments presented in~\cite{shotwisenoisyarxiv,cutandshootconf,bisicchia2026cut} indicate that shot-wise distribution can significantly reduce execution time and increase system resilience without imposing noticeable computational overhead.

The methodology also integrates naturally with other emerging paradigms. For instance, when combined with circuit cutting~\cite{cutqc, Peng}, shot-wise scheduling can be applied independently to each subcircuit, allowing large quantum workloads to be decomposed, distributed, and executed across a federation of QPUs. Ongoing research also explores noise-aware variants of the method~\cite{shotwisenoisyarxiv}, where knowledge of per-device error characteristics is used to allocate shots preferentially to higher-fidelity backends, thereby improving the quality of the aggregated result.

Despite its strengths, shot-wise distribution introduces new trade-offs. Submitting many small jobs increases the total number of interactions with quantum cloud services, potentially contributing to backend congestion and incurring higher scheduling complexity, a problem we are currently investigating. Moreover, result aggregation across devices with heterogeneous noise profiles can introduce bias if not appropriately accounted for. Nonetheless, prior evaluations~\cite{shotwisenoisyarxiv} suggest that these effects are manageable and, in some cases, recombining noisy fragments can even outperform execution on a single noisy device when supported by proper statistical correction.

In summary, shot-wise distribution offers a novel and practical mechanism for executing quantum circuits across distributed, heterogeneous infrastructures. It directly addresses some of the limitations of current quantum systems while aligning with the operational realities of cloud-based quantum computing. By offering both scalability and user-level control, this methodology may improve reliability, performance, and customizability in quantum software execution.

\subsection{Optimization Methods}

\paragraph{\rev{Mixed-Integer Linear Programming.}}
\rev{Mixed-integer linear programming (MILP) extends linear programming by
requiring some decision variables to take integer values, with binary
variables as an important special case. The objective function and
constraints remain linear, but the integrality restrictions make the
problem fundamentally different from a continuous linear program. In
the Quantum Broker, the shot-allocation variables are non-negative
integers and the backend-use variables are binary; therefore, the
Linear Solver implements a MILP rather than a continuous LP.}

\rev{Modern MILP solvers typically use branch-and-bound or branch-and-cut
methods and can certify global optimality with respect to the formulated
MILP when the optimisation terminates with an optimality certificate.
This guarantee applies only to the linearised optimisation model and
does not imply global optimality with respect to the original nonlinear
Profiler functions.}

\paragraph{Dual Annealing.}
Dual annealing~\cite{faber2006annealing} is a modern global optimisation algorithm that belongs to the family of stochastic and metaheuristic techniques. The core idea behind annealing algorithms is inspired by the physical process of heating a material and then gradually cooling it to remove defects and reach a stable, low-energy state. In optimisation, simulated annealing algorithms emulate this process by exploring the solution space in a manner that allows for occasional acceptance of non-optimal solutions, thereby enabling the search to escape local optima and continue towards better overall solutions.

Dual annealing enhances this approach by integrating simulated annealing with deterministic local search strategies, striking a balance between broad exploration and targeted exploitation of promising regions in the search space. The algorithm alternates between probabilistic jumps, where even worse solutions can be accepted with a certain probability, and focused, local improvement steps. This hybrid strategy helps dual annealing avoid becoming trapped in suboptimal solutions, which is a common challenge in complex, high-dimensional, or multimodal optimisation landscapes.

The principal strength of dual annealing is its robustness when dealing with objective functions that are nonlinear, non-convex, noisy, or not well understood analytically. It is suitable for a wide range of problems where traditional deterministic optimisation techniques may fail or become impractically slow. The trade-off, however, is that dual annealing is a probabilistic method: it does not guarantee finding the absolute global optimum and can require significantly more computational time, especially as the complexity of the problem increases. Nevertheless, its flexibility and ability to handle complex real-world conditions have made it a popular choice for challenging optimisation scenarios in science, engineering, and beyond.
\section{Related Work}
\label{sec:related}

The problem of selecting the most appropriate QPUs for executing a quantum circuit has received significant attention in the literature. Most existing approaches address QPU selection by defining and optimizing a set of metrics that may include not only the physical and logical characteristics of available quantum devices (such as qubit count, topology, or noise model), but also circuit-specific features (gate count, width, depth), and dynamic environmental factors such as queue waiting time, cost, and backend availability.

A prominent example is the \emph{Quantum API Gateway}~\cite{garcia2021quantum}, which provides a service for the automatic selection of an optimal QPU for each submitted quantum circuit. The selection process considers factors such as device architecture (e.g., gate-based vs. annealing) and circuit width, and enables users to prioritise speed or cost according to their application requirements.

The \emph{NISQ Analyzer}~\cite{nisqanalyzer} follows a similar strategy but goes further by acknowledging that multiple circuit implementations may exist for a single quantum algorithm. It selects the best circuit implementation from a repository of candidates using selection rules dependent on input data, and then determines the most suitable QPU-circuit pair according to criteria such as circuit width, depth, and SDK compatibility. The NISQ Analyzer ecosystem has been extended to include tools for compiler comparison~\cite{nisqanalyzercompar}, ML-based optimisation for discarding non-promising compilers and QPUs prior to compilation~\cite{nisqanalyzerml}, and for further compilation optimisation using machine learning~\cite{nisqanalyzeropt}.

Beyond these, several other approaches have investigated intelligent QPU selection and workload management. For instance,~\cite{adaptive} introduces an adaptive job scheduler that selects QPUs by jointly considering estimated output fidelity and queue waiting time. The framework in~\cite{qserverless} addresses the integration of quantum computing within enterprise cloud systems, using static attributes such as qubit count and queue length to select a suitable device. In a similar vein,~\cite{predictingml} presents a system for automatically predicting the best combination of QPUs, compilers, and compilation parameters to maximize execution fidelity for a given quantum circuit.

\new{
Recent work has also considered quantum execution from a broader resource-management and orchestration perspective. Ravi et al.~\cite{ravi2021adaptive} study adaptive job and resource management for quantum-cloud platforms, explicitly considering queueing time, fidelity trends across machines, user-level QoS guarantees, and calibration constraints. Their scheduler predicts execution fidelity from compiled circuit features and estimates queueing delay to select suitable machines under different system-load regimes. This work is closely related to ours because it frames quantum execution as a resource-management problem rather than as a purely algorithmic concern. However, its scheduling decision remains primarily at the level of assigning a complete quantum job to a machine, whereas the Quantum Broker reasons over the distribution of the shots of a single quantum computation across multiple heterogeneous backends subject to declarative QoS constraints.}

\new{
System-level abstractions for quantum resource management have also been proposed. QOS~\cite{giortamis2025qos} introduces a cloud operating system for quantum resources, exposing hardware-agnostic APIs and integrating mechanisms for error mitigation, fidelity estimation, multi-programming, and scheduling. Its objective is to balance quantum fidelity, resource utilisation, and waiting time in NISQ cloud platforms. QOS is complementary to our work: it focuses on operating-system support, cross-layer mechanisms, and provider-side scheduling of quantum jobs, while the Quantum Broker focuses on user-side, QoS-aware execution planning and shot-wise dispatch across multiple QPUs and providers.}

\new{
At the orchestration level, Qonductor~\cite{giortamis2025qonductor} proposes a cloud orchestrator for hybrid quantum-classical applications running on heterogeneous resources. It provides a hardware-agnostic API, estimates quantum and classical resource requirements, and schedules hybrid workloads by balancing user QoS objectives with cloud-operator resource-efficiency goals. This line of work is relevant because it addresses quantum-cloud orchestration under resource contention and heterogeneous backend performance. In contrast, the Quantum Broker targets a different granularity of orchestration: rather than scheduling complete hybrid applications, it optimises the allocation of measurement shots for a quantum workload according to explicit QoS constraints and objectives.}

\new{
Other approaches focus on provider-side QPU scheduling and resource allocation. QSRA~\cite{lu2024qsra} adapts classical CPU scheduling ideas to cloud-based quantum computing, incorporating qubit-quality and connectivity-aware allocation and merging multiple quantum programs to improve qubit utilisation and turnaround time. Similarly, Zheng et al.~\cite{zheng2024novel} propose a scheduling framework for multi-programming quantum computing in cloud environments. Their framework assigns tasks to quantum backends, considers fairness among users, employs multi-programming to improve throughput, and adaptively balances fidelity and user waiting time. These works further confirm the importance of scheduling and resource management in scarce quantum-cloud infrastructures. Nevertheless, they primarily address multi-user task scheduling, intra-QPU allocation, and backend utilisation. By contrast, the Quantum Broker operates at the application-execution level: given a single submitted quantum computation, it exploits shot independence to distribute its execution across multiple heterogeneous backends while satisfying user-defined QoS requirements.
}

Despite these advances, existing proposals predominantly focus on identifying a single best QPU (or QPU-circuit pair) for the entire quantum workload, and then executing all measurement shots on that device. By contrast, our approach departs from this paradigm by leveraging the statistical independence of quantum measurement shots. We employ the \emph{shot-wise distribution} methodology~\cite{shotwisenoisyarxiv,shotwiseicsoc}, wherein the shots (even those from a single circuit) are distributed across multiple, heterogeneous QPUs, thereby enabling parallelism, increased resilience, and adaptive load balancing.

To the best of our knowledge, this is the first general-purpose framework to realise QoS aware, shot-wise distribution for quantum computations. 
\section{Architecture \& Implementation}
\label{sec:broker}

This section presents the architecture of the \emph{Quantum Broker}\footnote{Quantum Broker is open-source and freely available at: \url{https://github.com/di-unipi-socc/TheQuantumBroker}.}, designed as a general, modular, and portable solution for managing QoS-aware quantum execution across heterogeneous, multi-provider environments. We begin by discussing the high-level system architecture and interaction patterns among the user, the system’s core components, and external QPUs. We then detail the execution flow, followed by focused subsections describing each major component.

\subsection{Overview and Interaction Model}

At its core, the Quantum Broker is conceived as a mediation layer between quantum software users and a dynamic pool of diverse quantum backends. The architecture is deliberately modular, supporting extensibility and clear separation of concerns. Figure~\ref{fig:arch} illustrates the main components and their interactions.

\Figure[!ht][width=0.9\textwidth]{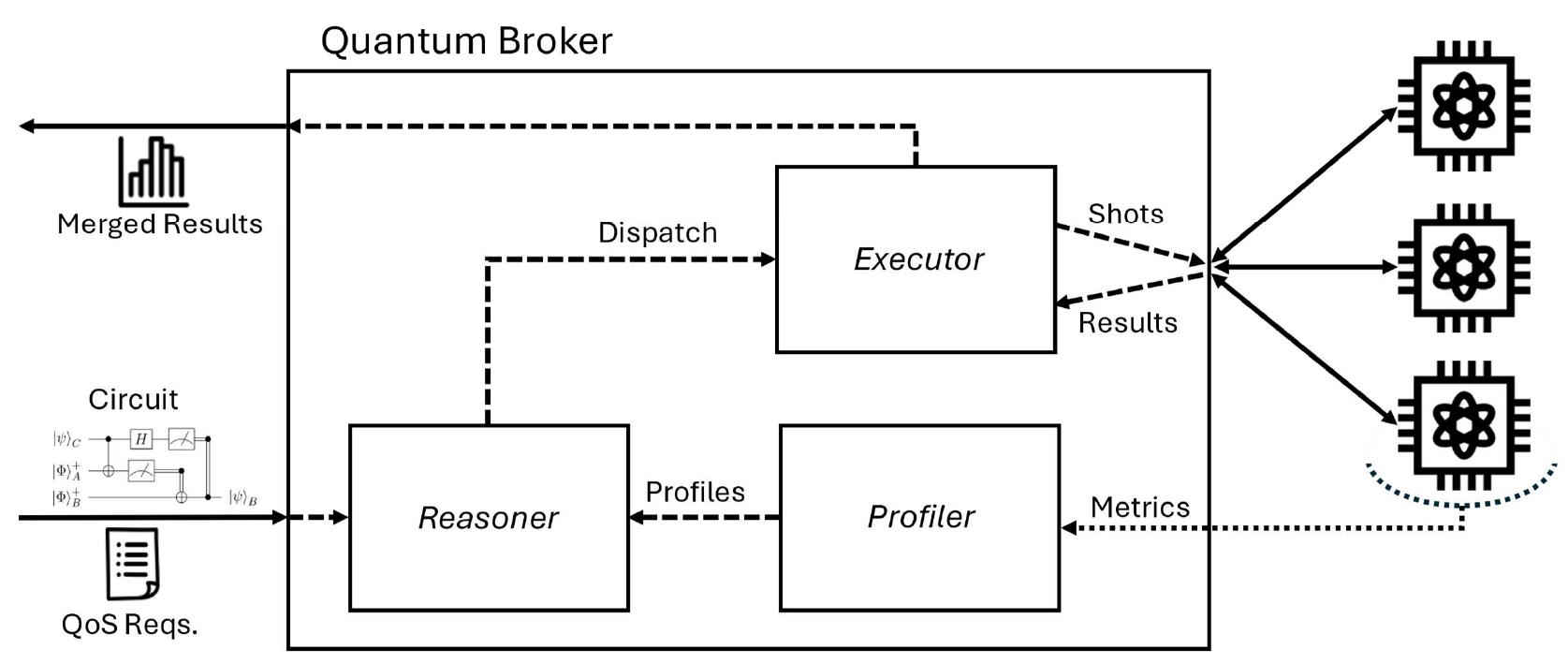}
{High-level architecture of the Quantum Broker.\label{fig:arch}}

\paragraph{User Interaction.} The user interacts with the Quantum Broker through a high-level, provider-agnostic API. The user submits a quantum circuit, together with a set of execution requirements and QoS constraints. These requirements can express both hard constraints (e.g., minimum fidelity, device region, budget cap) and optimization objectives (e.g., minimize total runtime, maximize throughput, balance shot allocation across devices). Requirements are specified using a portable, declarative format, making them easily extensible and automatable.

\paragraph{Component Overview.} The architecture is composed of three primary, loosely-coupled components:

\begin{enumerate}
    \item \textbf{Profiler}: Responsible for discovering available QPUs, extracting both static and dynamic device information (e.g., qubit count, error rates, queue depth, access pricing), and providing a unified abstraction over heterogeneous provider APIs and telemetry endpoints.
    \item \textbf{Reasoner}: Acts as the "decision engine” of the broker. It receives the user’s workload and requirements, accesses QPU metadata from the Profiler, and computes an optimal (or near-optimal) execution plan. The Reasoner determines how to allocate shots, select devices, and assign workloads according to user objectives and current system state.
    \item \textbf{Executor}: Implements the actual interaction with quantum backends. It orchestrates the distributed execution of the workload by submitting jobs to selected QPUs, monitors their progress, collects results, and performs result aggregation as required.
\end{enumerate}

The modularity of this design enables each component to be instantiated or extended independently. For example, the Reasoner may use pluggable policies (e.g., greedy, adaptive, or ML-based), while the Profiler can incorporate new benchmarking techniques as the quantum ecosystem evolves.


\subsubsection*{General Execution Flow}

The typical execution flow in the Quantum Broker can be summarized as follows:

\begin{enumerate}
    \item \textbf{Request Submission}: The user submits a quantum workload (i.e., a circuit and a required number of shots) and a QoS specification to the Quantum Broker API.
    \item \textbf{Profiling}: The Profiler discovers all available QPUs and collects up-to-date device metadata from multiple providers. It may use a combination of static manifests, provider APIs, web scraping, and live benchmarking.
    \item \textbf{Planning}: The Reasoner receives the workload, user requirements, and QPU profiles. It generates a candidate execution plan, which defines (i) how many shots to assign to each QPU, (ii) the order and grouping of tasks, and (iii) fallback or adaptation policies.
    \item \textbf{Execution}: The Executor dispatches the workload in accordance with the execution plan. Jobs are submitted to the appropriate QPUs, potentially in parallel and across multiple providers. The Executor monitors job status, adapts to failures (e.g., device downtime), and retrieves partial or final results as they become available.
    \item \textbf{Aggregation and Post-Processing}: Results from different QPUs are aggregated by the Executor according to user-defined or default policies (e.g., weighted average, fidelity-aware merging~\cite{bisicchia2025quantumexecutor,shotwisenoisyarxiv}). The final results are returned to the user, together with metadata such as cost, execution time, and device utilization.
\end{enumerate}

\paragraph{Design Motivations.}
Several key motivations underlie this architecture:

\begin{itemize}
    \item \textbf{User-Centric QoS}: By providing a portable requirements interface, the system empowers users to articulate diverse, high-level objectives, beyond what any single provider or backend currently supports.
    \item \textbf{Decoupling and Portability}: By strictly separating user intent, decision logic, and backend interaction, the Quantum Broker remains portable across evolving quantum and classical environments.
    \item \textbf{Dynamic Adaptation}: The explicit profiling and reasoning layers allow the broker to adapt in real-time to backend variability, device failures, or sudden changes in provider offerings.
    \item \textbf{Extensibility and Experimentation}: Each major component supports pluggable implementations and policies. This enables research and practical experimentation with alternative decision strategies (e.g., cost-aware, latency-aware, green computing), alternative profiling methods, or advanced aggregation techniques.
    \item \textbf{Reproducibility and Transparency}: All decisions are made explicit in the execution plan, supporting experiment reproducibility and post-hoc analysis.
\end{itemize}

For clarity and completeness, the next subsections describe each major component in detail: their responsibilities, inner functioning, interfaces, and key design choices.

\subsection{Executor}

The \emph{Executor} is a core component of the Quantum Broker framework, designed to address a fundamental need in modern quantum software engineering: enabling portable, robust, and scalable execution of quantum experiments across a rapidly evolving landscape of quantum hardware and cloud providers.


The Executor~\cite{bisicchia2025quantumexecutor} is conceived as a backend-agnostic engine that abstracts away the complexity of individual quantum providers and SDKs. Its fundamental purpose is to serve as an orchestration layer between the user's intent (expressed as quantum circuits and execution requirements) and the underlying pool of available quantum and classical resources.

When a quantum experiment is submitted, the Executor performs the following core functions:
\begin{itemize}
    \item \textbf{Unified Experiment Submission:} Users define their experiments and requirements using a declarative, provider-independent interface. The Executor accepts these experiments without requiring adaptation to specific backend APIs or SDKs.
    \item \textbf{Orchestration and Execution:} The Executor handles the distribution of jobs across the chosen backends. It supports both parallel and asynchronous execution, allowing workloads to be dispatched to multiple QPUs and simulators simultaneously, according to user requirements or system policies.
    \item \textbf{Result Collection and Aggregation:} As quantum jobs are executed, results are collected in real time. The Executor provides mechanisms for aggregating and merging results, ensuring that outputs from heterogeneous devices are presented in a coherent and unified manner.
    \item \textbf{Extensibility and Portability:} By decoupling experiment logic from backend specifics, the Executor ensures that users can repurpose experiments as the quantum ecosystem evolves, and can integrate new hardware or simulators with minimal friction.
\end{itemize}

The \emph{input} to the Executor is a \textbf{dispatch}, a declarative, hierarchical data structure that describes the full execution plan. The dispatch explicitly encodes:

\begin{itemize}
\item \textbf{Provider Selection:} Which quantum/cloud provider should handle each job batch.
\item \textbf{Backend Assignment:} Which specific device(s) or simulators are to be used within each provider.
\item \textbf{Task Specification:} For each backend, a list of execution tasks, each defined by a quantum circuit (potentially in various formats) and an associated shot count.
\end{itemize}

Formally, a dispatch is represented as a nested dictionary structure:
\begin{modelview}[label=dispatch]{An example of Quantum Dispatch}
{\fontsize{9}{9}\selectfont
\begin{verbatim}
{
    "provider_1": {
        "backend_A": [
            {
                "circuit": circuit_1, 
                "shots": 1024
            },
            {
                "circuit": circuit_2, 
                "shots": 2000
            },
        ],
        "backend_B": [
            {
                "circuit": circuit_3, 
                "shots": 256
            },
        ],
    },
    "provider_2": {
        "backend_C": [
            {
                "circuit": circuit_4, 
                "shots": 6000
            },
        ]
    }
}
\end{verbatim}
}
\end{modelview}
Each task is executed on the designated backend with the specified number of shots. Circuits may be specified in different programming languages or formats (e.g., Qiskit, Cirq, PennyLane, OpenQASM\footnote{See \url{https://github.com/GBisi/quantum-executor} for the full list}), and the Executor handles necessary conversions and transpilation internally.

Upon receiving a dispatch, the Executor performs the following core functions:

\begin{itemize}
\item \textbf{Unified Experiment Submission:} Accepts the dispatch via a provider-agnostic interface, requiring no user adaptation for provider- or backend-specific APIs.
\item \textbf{Orchestration and Execution:} Distributes jobs as described in the dispatch across all specified backends. Supports parallel and asynchronous execution, enabling efficient use of multi-provider, multi-device, and hybrid cloud resources.
\item \textbf{Monitoring and Live Result Access:} Tracks execution status for each job, supporting partial result retrieval and real-time monitoring. Jobs that finish earlier can be inspected without waiting for global completion.
\item \textbf{Result Collection and Aggregation:} Aggregates results from all backends. Provides hooks for user-defined or policy-driven merging, enabling advanced ensemble or statistical analyses across heterogeneous devices.
\item \textbf{Extensibility and Portability:} Decouples experiment logic and dispatch structure from backend specifics. New providers, devices, or circuit formats can be integrated with minimal friction by extending provider adapters.
\end{itemize}

The Executor is designed around the following architectural principles:
\begin{itemize}
\item \textbf{Backend-agnosticism:} Supports all major providers (e.g., IBM Qiskit, IonQ, AWS Braket, Azure Quantum, local simulators\footnote{See \url{https://github.com/GBisi/quantum-executor} for the full list}) without requiring backend-specific logic in user code.
\item \textbf{Language and Format Independence:} Accepts circuits written in a wide variety of quantum programming languages and representations, handling all necessary conversion.
\item \textbf{Custom Execution Policies:} Supports user-defined split (dispatch) and merge (aggregation) policies, enabling optimization for cost, runtime, noise level, or other constraints.
\item \textbf{Parallelism and Asynchrony:} Enables asynchronous, non-blocking, and parallel execution across distributed quantum and classical resources.
\item \textbf{Declarative Workflow:} Moving a workflow to a different provider or backend requires only a change in the dispatch, not in core code, supporting portable and reproducible research and production pipelines.
\end{itemize}

In summary, the Executor receives a complete, explicit dispatch describing the mapping from circuits to providers and backends, and executes the plan in a unified, extensible, and backend-agnostic fashion. This architectural approach supports rapid experimentation, system scalability, and seamless evolution as the quantum computing ecosystem advances.

For a detailed architectural discussion, please see~\cite{bisicchia2025quantumexecutor} and the open-source repository\footnote{\url{https://github.com/GBisi/quantum-executor}}.

\subsection{Profiler}

The \emph{Profiler} is a core abstraction of the Quantum Broker architecture. Its purpose is to provide a unified, actionable, and up-to-date view of all accessible QPUs and their operating environments. The Profiler enables dynamic, QoS-aware orchestration by supplying the Reasoner with device metadata, performance models, and execution cost predictions. Crucially, it abstracts away provider heterogeneity, delivering a standardized interface for QPU discovery and characterization across vendors and platforms.

Architecturally, the Profiler is conceived as a modular component whose responsibility is to collect, consolidate, and expose both static (e.g., qubit count, topology, provider name) and dynamic (e.g., error rates, queue depths, live performance metrics) properties of all candidate backends. It mediates between the Broker and the diverse quantum/cloud infrastructure, insulating higher layers from API fragmentation and reporting inconsistencies.

The Profiler provides:
\begin{itemize}
\item Uniform schema for device metadata and performance characteristics.
\item Pluggable strategies for device discovery and benchmarking.
\item Extensible interface for integrating future metrics and data sources.
\item Predictive models for execution time, cost, and expected fidelity.
\end{itemize}

This design enables the Profiler to serve as a foundation for reproducible planning and adaptive dispatch, regardless of how backend data is gathered.

The Profiler is explicitly designed for modularity: its interface allows for multiple, interchangeable implementations. For example, one implementation may rely exclusively on provider-reported metadata and static device manifests. Another may incorporate live telemetry feeds or advanced benchmarking protocols. Still others could use external databases or crowd-sourced reputation scores.

This architectural decoupling supports experimentation with novel profiling techniques as the quantum ecosystem evolves. It also permits the system to swap profiling backends at runtime, accommodating user or organizational policy, or adapting to new hardware without requiring system-wide changes.

\paragraph{Prototype Implementation: Empirical Benchmarking.}
To illustrate the potential of empirical, experiment-driven profiling, we developed a prototype Profiler that actively benchmarks each QPU using synthetic quantum circuits. The key objective is to produce fine-grained, empirically validated models of error, performance, and cost, which are then used to predict workload outcomes.

The protocol proceeds as follows:

\begin{enumerate}
\item \textbf{Parameterization.} A configurable fraction of available qubits is selected, and benchmarking parameters (max circuit depth, layer size, shot count) are set to balance accuracy and experimental overhead.
\item \textbf{Random Circuit Generation.} For each circuit depth (up to the max circuit depth), a random quantum circuit is synthesized over the selected qubits, using fixed seeds for reproducibility.
\item \textbf{On-Device vs. Simulated Execution.} Each circuit is executed both on the target QPU (via its native provider) and a noise-free classical simulator. Execution times are recorded for each run.
\item \textbf{Fidelity Assessment.} \rev{Throughout this paper, we use the term \emph{distributional fidelity} specifically for output-distribution similarity measured with the squared Bhattacharyya coefficient. This quantity is distinct from the synthetic reliability score used later in the controlled evaluation. Given the empirical output distribution obtained from the backend, \(p_{\mathrm{hw}}\), and the reference distribution obtained from a noise-free simulator, \(p_{\mathrm{ideal}}\), the empirical Profiler defines}

\[
F_{\mathrm{BC}}(p_{\mathrm{hw}},p_{\mathrm{ideal}})
=
\left(
\sum_{x \in \mathcal{X}}
\sqrt{
p_{\mathrm{hw}}(x)\,p_{\mathrm{ideal}}(x)
}
\right)^2 ,
\]

\rev{where \(\mathcal{X}\) is the union of the bitstrings observed in either distribution. We denote this quantity by \(F_{\mathrm{BC}}\) and refer to it as \emph{empirical distributional fidelity}. It lies in \([0,1]\), with \(1\) indicating identical empirical distributions and lower values indicating increasing divergence from the ideal noiseless behaviour. It is not a quantum-state fidelity between density operators.}
\item \textbf{Decay and Timing Modelling.} The observed fidelity values across layers are fit to an exponential decay model, parameterized by a per-layer decay rate $\alpha$. Execution times yield average per-experiment duration. \new{For each benchmark circuit depth, the Profiler computes \(F_{\mathrm{BC}}\) between the backend output distribution and the ideal simulator output distribution. The resulting sequence of fidelity values is then fitted with an exponential decay model. If \(n\) denotes the number of benchmark layers, the fitted model is}

\[
F_{\mathrm{BC}}(n) \approx \alpha^n ,
\]

\new{where \(\alpha \in [0,1]\) captures the average fidelity retention per benchmark layer. Since each benchmark layer has depth \(d_{\mathrm{layer}}\), the per-depth fidelity retention is estimated as}

\[
\alpha_{\mathrm{depth}} = \alpha^{1/d_{\mathrm{layer}}},
\]

\new{and the corresponding per-depth error parameter stored in the backend profile is}

\[
\epsilon_{\mathrm{depth}}
=
1-\alpha_{\mathrm{depth}} .
\]

\new{This is the quantity reported as \texttt{error\_per\_depth} in the cached QPU profile.}
\item \textbf{Profile Extraction.} The profile summarizes:
\begin{itemize}
\item \texttt{error\_per\_depth}: The per-layer error rate ($1 - \alpha^{1/\texttt{layer\_depth}}$).
\item \texttt{average\_time}: Mean execution time per experiment.
\item Benchmarking parameters: effective qubits, shot count, total depth, randomization seed.
\end{itemize}
\end{enumerate}

\noindent
An example of the raw experimental data underlying profile construction is shown in Listing~\ref{actualprofile}.

\begin{modelview}[label=actualprofile]{QPU Experimental Data}
{\fontsize{9}{9}\selectfont
\begin{verbatim}
{
    "error_per_depth": 0.0034,
    "average_time": 49.384847,
    "effective_qubits": 10,
    "shots": 4096,
    "max_layers": 10,
    "max_depth": 50,
    "layer_depth": 5,
    "qubits_fraction": 1.0,
    "accuracy": 100,
    "max_qubits": 10,
    "seed": 42,
    "shots_alpha": 4,
    "total_depth": 50
}
\end{verbatim}
}
\end{modelview}


\new{\paragraph{Profile construction and dispatch-time elaboration.} The empirical benchmarking procedure described above is not repeated for every user dispatch. Instead, it is executed offline, periodically, or on demand when a backend profile must be refreshed. Its output is a cached backend profile containing backend-level parameters such as \texttt{error\_per\_depth}, \texttt{average\_time}, benchmarking configuration, and related metadata.}

\new{At dispatch time, the Profiler performs only a lightweight elaboration step: it combines the most recent cached profile for each backend with the user-submitted circuit and requested number of shots. This elaboration extracts circuit-specific information, such as logical depth, and uses the cached profile to estimate the QoS quantities needed by the Reasoner, including expected fidelity, execution time, cost, and waiting time.}

Given a user-submitted quantum circuit and the number of required shots, the elaboration process involves the following key steps:

\begin{enumerate}
\item \textbf{Depth Extraction.} The input circuit is transpiled to a standard format (e.g., Qiskit) and its logical depth is computed. This depth is then used as the basis for fidelity and time estimation.
\item \new{\textbf{Fidelity Estimation.}}
\rev{At dispatch time, the empirical distributional fidelity is not recomputed by rerunning the benchmark circuits. Instead, the Profiler uses the cached per-depth error parameter, \(\epsilon_{\mathrm{depth}}\), obtained by fitting the measured \(F_{\mathrm{BC}}\) values during profile construction. For a user-submitted circuit with logical depth \(d\), it predicts a depth-dependent distributional-fidelity proxy through the exponential decay model}

\[
F_{\mathrm{dispatch}}
=
\max\left(
0,
\min\left(
1,
(1-\epsilon_{\mathrm{depth}})^d
\right)
\right).
\]

\rev{We denote this dispatch-time prediction by \(F_{\mathrm{dispatch}}\). It is a backend-specific, depth-dependent proxy derived from empirical distributional similarity, not a newly observed fidelity value and not a quantum-state fidelity. Thus, the empirical profiling path uses two related but distinct quantities: measured distributional fidelity \(F_{\mathrm{BC}}\) during benchmarking and its fitted dispatch-time prediction \(F_{\mathrm{dispatch}}\).}

\item \textbf{Execution Time Estimation.} Execution time is computed differently depending on the provider:
\begin{itemize}
    \item For \textbf{IBM} backends, the profiler uses Qiskit's built-in duration estimator on the transpiled circuit and scales the result by the number of shots.
    \item For \textbf{IonQ} devices, a heuristic model is used, based on the ratio of 1-qubit and 2-qubit gates, scaled by their respective durations (150µs and 600µs), plus readout overhead. This is multiplied by the shot count and normalized to seconds.
    \item For \textbf{local simulators}, execution time is derived empirically using the average time per total depth collected during profiling. This is scaled proportionally to the circuit depth and number of shots.
\end{itemize}

\item \textbf{Cost Estimation.} Cost is similarly backend-specific:
\begin{itemize}
    \item For IBM, cost is approximated as a linear function of execution time, scaled by a fixed rate (e.g., 1.6 cost units per second).
    \item For IonQ, a constant base price is combined with a shot-dependent term (e.g., \$0.3 base + \$0.03/shot or \$0.08/shot for specific devices).
    \item Simulators are typically assumed to be free or at negligible cost.
\end{itemize}

\item \textbf{Waiting Time Estimation.} If not already known from the cached profile, waiting time is inferred by subtracting estimated execution time from the previously observed average total duration. For simulators or backends with on-demand access, this may be zero or null.

\end{enumerate}

The result is a consistent, per-backend estimate of the key QoS parameters (such as in Listing~\ref{profiler}):

\begin{modelview}[label=profiler]{A QPU Profile}
{\fontsize{9}{9}\selectfont
\begin{verbatim}
{
    "cost": 4.2,
    "execution_time": 0.08,
    "waiting_time": 112,
    "fidelity": 0.9996
}
\end{verbatim}
}
\end{modelview}

\begin{itemize}
\item \texttt{fidelity}: \rev{a profiler-provided backend-quality value. For the empirical Profiler, this field is populated with the predicted distributional fidelity \(F_{\mathrm{dispatch}}\).}
\item \texttt{execution\_time}: expected time in seconds to complete the full job (including scaling by shots).
\item \texttt{cost}: total estimated cost of executing the job.
\item \texttt{waiting\_time}: delay due to queueing, inferred from past empirical data.
\end{itemize}

\medskip

This elaboration enables the Reasoner to reason over realistic QoS metrics tailored to the specific circuit and backend, rather than relying on static device specifications. Importantly, because the estimation models are based on experimentally observed behaviour and backend-specific heuristics, they can capture nontrivial performance patterns and provider idiosyncrasies.

In summary, the Profiler is an architecturally modular, implementation-agnostic component, designed to adapt to both the diversity and volatility of the quantum landscape. Our prototype aims to illustrate the power of empirical, experiment-driven profiling, but the framework supports the integration of alternative and more sophisticated strategies as new techniques, provider APIs, or community standards emerge. In all cases, the Profiler’s core mission is to deliver rigorous, real-time situational awareness as the basis for reproducible and adaptive quantum workload management.

\subsection{QoS Modelling}

The efficient deployment of quantum circuits in heterogeneous quantum computing environments requires precise specification of Quality of Service (QoS) constraints. Given a quantum circuit and a set of candidate quantum backends (with different capabilities and queue states), it is essential to define how execution resources should be allocated to meet cost, time, and fidelity objectives. To this end, we introduce a dedicated, declarative \textit{\qoslanguage} (\qoslanguageacr), designed to compactly express constraints and optimisation objectives that guide the dispatch of quantum workloads.

\qoslanguageacr enables users to formulate constraints and objectives using a set of domain-specific variables, arithmetic and aggregation operators, and backend indexing. Each constraint defines a property that a valid dispatch must satisfy, while the objective function determines the target for optimisation (such as minimising cost or total execution time).

The core syntax of \qoslanguageacr is described by the following EBNF grammar shown in Listing~\ref{ebnf:qos}.
It is important to note that a set of implicit constraints is always automatically enforced by the system, and therefore does not need to be specified by the user. These include: 
\begin{itemize}
    \item Shot assignments are always non-negative integers. 
    \item The total number of shots assigned across all backends must exactly equal the required total.
    \item At least one backend must be used.
    \item A backend is marked as used if and only if it is allocated at least one shot.
\end{itemize}

\noindent
These implicit constraints provide the fundamental consistency required for any feasible quantum workload deployment.

\begin{ebnfbox}[title=Listing~\thetcbcounter:~\qoslanguage Grammar (EBNF)]{ebnf:qos}
{\fontsize{9}{9}\selectfont
\begin{verbatim}
statement      ::= constraint | objective

constraint     ::= expression 
                   comparator 
                   expression

objective      ::= expression

expression     ::= term { operator term }

term           ::= aggregate
                | idx_variable
                | variable
                | number
                | "(" expression ")"

aggregate      ::= ("sum" | "max" | "min") 
                   "(" expression ")"

idx_variable   ::= variable "[" index "]"

variable       ::= "shots"
                | "used"
                | "cost"
                | "execution_time"
                | "waiting_time"
                | "fidelity"
                | "total_shots"
                | weight_name

index          ::= integer
                | backend_name

operator       ::= "+" | "-" | "*" |
                   "/" | "**"

comparator     ::= "<" | ">" | "<=" | 
                   ">=" | "==" | "!="
\end{verbatim}
}
\end{ebnfbox}

At the heart of the \qoslanguageacr lies a set of domain-specific variables calculated that reflect the essential quantities and qualities of quantum deployment. These include, for instance, the number of shots assigned to each backend, binary usage indicators (\texttt{used}), estimated cost, expected execution and waiting times, and the achieved fidelity for each backend. Importantly, each of these variables may be indexed, allowing precise referencing either by backend position (an integer index) or by backend name (a quoted string). This dual-mode indexing facilitates robust and readable constraint specification; for example, one can require that at least 100 shots be assigned to the backend named ``fake\_torino'' by writing \verb|shots['fake_torino'] >= 100|.

A distinguishing aspect of the language is its support for aggregation functions, namely, \texttt{sum}, \texttt{max}, and \texttt{min}. These aggregates enable concise expressions over the set of all backends, supporting global constraints such as limiting the total cost (\texttt{sum(cost) <= 5000}), enforcing a minimum fidelity across all used devices (\texttt{min(fidelity) >= 0.9}), or bounding the maximum expected completion time (\texttt{max(execution\_time + waiting\_time) <= 1200}). Aggregation, together with the flexible indexing, allows both global and per-backend properties to be captured in a uniform way.

The language adopts standard arithmetic operators, including addition, subtraction, multiplication, division, and exponentiation, to build composite expressions. Furthermore, any \verb|weight_name| can be used \new{to introduce} user-defined weights which makes it possible to encode customised optimisation preferences directly in the objective function. A typical objective might minimise a weighted sum of total cost and the maximum turnaround time, as in \texttt{cost\_weight * sum(cost) + time\_weight * max(execution\_time + waiting\_time)}. Such expressions reflect the nuanced trade-offs often encountered in quantum workload management, where cost, latency, and quality must be balanced.

Constraints in \qoslanguageacr are always formulated as comparisons between two expressions using logical comparators (\texttt{<}, \texttt{>}, \texttt{<=}, \texttt{>=}, \texttt{==}, \texttt{!=}). This structure allows for a wide variety of requirements to be captured---from simple limits on individual variables to complex relationships across multiple aggregated quantities. The objective, in turn, is any valid expression constructed from the available variables, aggregates, and operators; it defines the goal to be optimised, such as minimising cost, maximising fidelity, or achieving a particular weighted combination of both.

Taken together, these features give \qoslanguageacr a unique combination of expressiveness and usability. By abstracting the details of backend heterogeneity and providing a powerful constraint/objective formalism, it enables users and systems to precisely state their requirements for quantum circuit deployment. This clarity and structure are crucial for enabling automated solvers to produce optimal or near-optimal resource allocations that respect all specified QoS criteria.

\subsubsection*{Illustrative Scenarios in \qoslanguageacr}
\label{sec:qml}

To illustrate the expressive power and flexibility of \qoslanguageacr, we present three representative quantum deployment scenarios, each designed to encode realistic resource and performance requirements. For each scenario, we explain the intent behind the constraints and objectives, present the corresponding \qoslanguageacr encoding, and briefly discuss the practical implications of the specified model.

\new{
The scenarios used in this work are intentionally simple and controlled. Their purpose is not to exhaustively
represent all possible real-world quantum deployment policies, but to isolate the main decision dimensions that a
QoS-aware quantum broker must handle: cost, execution time, waiting time, fidelity, and backend-specific allocation
requirements. By keeping each scenario compact, we make the effect of each constraint and objective interpretable,
reproducible, and directly comparable across solvers and baselines.}

\new{This choice is particularly important because the Quantum Broker is evaluated as an optimisation and dispatch
framework, rather than as a domain-specific scheduler for a single application. More complex policies can be
constructed compositionally from the same QQSpec primitives used in these scenarios: arithmetic expressions,
aggregations over backends, per-backend indexing, hard constraints, and weighted objectives. Therefore, the selected
scenarios act as representative test cases for the expressiveness of QQSpec and for the behaviour of the Reasoner
under different classes of QoS requirements.}

\new{The three scenarios were chosen to cover complementary dispatch regimes. Scenario~1 combines multiple
simultaneous constraints with a balanced cost--time objective, stressing the Broker under potentially conflicting
requirements. Scenario~2 studies budget utilisation under fidelity and time constraints, testing whether the solver
can exploit available resources without violating QoS bounds. Scenario~3 introduces an explicit backend assignment
constraint, representing cases where a user or organisation requires that a specific device be used, for example for
benchmarking, calibration, policy, or availability reasons. Together, these scenarios provide a compact but meaningful
basis for evaluating whether QoS-aware shot-wise distribution improves over single-backend and naive multi-backend
dispatch strategies.}

\paragraph{Scenario 1: Balanced optimisation with cost, fidelity, and throughput guarantees.}

In this scenario, the user wants to split work among available quantum backends, ensuring that each backend receives at least 300 shots to have reliable results. All used backends must maintain a fidelity of 0.9. The total cost of the dispatch must satisfy the budget of 800 unit cost, and overall time must be under 2000 seconds. Since the quantum execution is parallelised, the performance is dictated by the slowest backend, so the objective is to minimise the maximum total time (execution time plus waiting time) required by any backend.
The objective is to find a balance between fast execution and low cost, tuning this trade-off equally with a linear combination of the two quantities to minimise with the same weight (0.5)\footnote{\rev{For Scenario~1, we retain the original scalarisation
$
J = 0.5\,T + 0.5\,C
$
used throughout the controlled benchmark. The coefficients are
therefore treated as fixed scaling factors rather than as
dimensionless preference weights. The synthetic backend profiles were
constructed such that the cost and completion-time terms entering the
objective have comparable numerical magnitudes over the evaluated
instances, preventing either term from dominating the scalarisation
solely because of its numerical scale. This choice is specific to the
controlled experiment and is not intended as a transferable weighting
scheme for production deployments.}}. 

\rev{The objective is a scalarisation of cost and completion time. Because these quantities have different physical units and may have different numerical scales, the coefficients in such a scalarisation must not be interpreted as dimensionless preference weights unless the QoS terms have first been normalised. In general, we recommend using reference values, e.g., $\widetilde{C}=C/C_{\mathrm{ref}}$ and $\widetilde{T}=T/T_{\mathrm{ref}}$, and optimising $w_C\widetilde{C}+w_T\widetilde{T}$ with $w_C+w_T=1$. QQSpec does not perform this normalisation automatically: the user must either normalise the terms explicitly or choose scaling coefficients with the appropriate reciprocal units. In this controlled synthetic benchmark, cost and time were parameterised to occupy comparable numerical ranges, and the coefficients 0.5/0.5 are used only as a simple experimental scalarisation; they should not be transferred unchanged to deployments with different units or ranges.}

\begin{modelview}[label=model:scenario1]{Scenario 1}
{\fontsize{9}{9}\selectfont
\begin{verbatim}
weights:
    time_weight = 0.5
    cost_weight = 0.5

constraints:
    sum(cost) <= 800
    min(shots) >= 300
    min(fidelity) >= 0.9
    max(execution_time 
        + waiting_time) <= 2000

objective:
    time_weight * 
    max(execution_time + waiting_time) + 
    cost_weight * sum(cost)
\end{verbatim}
}
\end{modelview}

\noindent
This scenario is expressed with \qoslanguageacr as displayed in Listing~\ref{model:scenario1}, and it shows how to express an objective with a trade-off between different variables and how constraints can have minimum and maximum bounds for all the backends.

\paragraph{Scenario 2: Maximising budget utilisation under fidelity and time constraints.}

\rev{Scenario 2 intentionally uses an unusual maximisation objective to exercise QQSpec's ability to express a maximisation problem under hard QoS constraints. The total shot budget remains fixed at 4096, so maximising expenditure does not increase the amount of quantum work; rather, it favours dispatches that use more of a capped budget while still satisfying the reliability and time requirements. A practical interpretation is a use-it-or-lose-it allocation of prepaid quantum-cloud credits, or an organisational policy that prefers consuming an already committed budget on premium resources before it expires. All used backends must provide a synthetic reliability score of at least 0.85, the overall time (execution plus waiting time) must not exceed 1700 seconds, and total cost must remain below 750 units. The maximisation is encoded by minimising $-\sum \mathrm{cost}$.}

\begin{modelview}[label=model:scenario2]{Scenario 2}
{\fontsize{9}{9}\selectfont
\begin{verbatim}
constraints:
    max(execution_time 
        + waiting_time) <= 1700
    min(fidelity) >= 0.85
    sum(cost) <= 750
    
objective:
    -sum(cost)
\end{verbatim}
}
\end{modelview}

\noindent
Listing~\ref{model:scenario2} illustrates how to use \qoslanguageacr to express this scenario with a maximisation objective (via a negative sign) under combined budget, quality, and time requirements.

\vspace{2ex}

\paragraph{Scenario 3: Targeted backend usage with global fidelity and cost constraints.}

In this scenario, the user requires that a particular backend, \texttt{"fake\_torino"}, is assigned at least 700 shots, perhaps due to calibration needs or specific benchmarking. The sum of the costs for all selected backends must stay within a budget of 700 units. The objective is to minimise the overall time, and for this reason, the expression \texttt{max(execution\_time + waiting\_time)} is used to capture the longest execution time among the selected backends.

\begin{modelview}[label=model:scenario3]{Scenario 3}
{\fontsize{9}{9}\selectfont
\begin{verbatim}
constraints:
    shots['fake_torino'] >= 700
    sum(cost) <= 700

objective:
    max(execution_time + waiting_time)
\end{verbatim}
}
\end{modelview}

\noindent
This scenario's \qoslanguageacr is showcased in Listing~\ref{model:scenario3} to exhibit how to express the targeting of a specific backend with an assignment constraint, while still imposing global requirements, e.g., on the budget. The model also illustrates how to set the objective to optimise for the best possible completion time across all used resources.

\medskip
\noindent
These scenarios highlight how \qoslanguageacr supports a wide variety of real-world quantum deployment policies, enabling users to encode nuanced requirements for workload distribution, performance, cost, and backend selection with concise and transparent expressions.

\subsection{Reasoner}

The \emph{Reasoner} is the intelligent core of the system, tasked with transforming the user's high-level Quality of Service requirements into a concrete resource allocation plan for quantum circuit execution (i.e., a \textit{Dispatch}). Upon receiving as input a QoS model, a quantum circuit, and the total number of shots, the Reasoner parses the constraints and objective, and prepares the problem for optimisation. \rev{In our implementation, depending on user preference, the Reasoner can employ one of two solver backends: (i) a mixed-integer linear programming (MILP) solver for fast optimisation of a linearised QoS surrogate (the ``Linear Solver''), or (ii) a dual annealing solver for profile-driven optimisation in the presence of nonlinear effects.} After instantiating the selected solver and running the optimisation, the Reasoner produces as output a dispatch plan specifying the shot allocation across backends, if a feasible solution satisfying all requirements is found.

Figure~\ref{fig:reasonerarch} illustrates the architecture and workflow of the Reasoner component.

\Figure[!ht][width=0.6\textwidth]{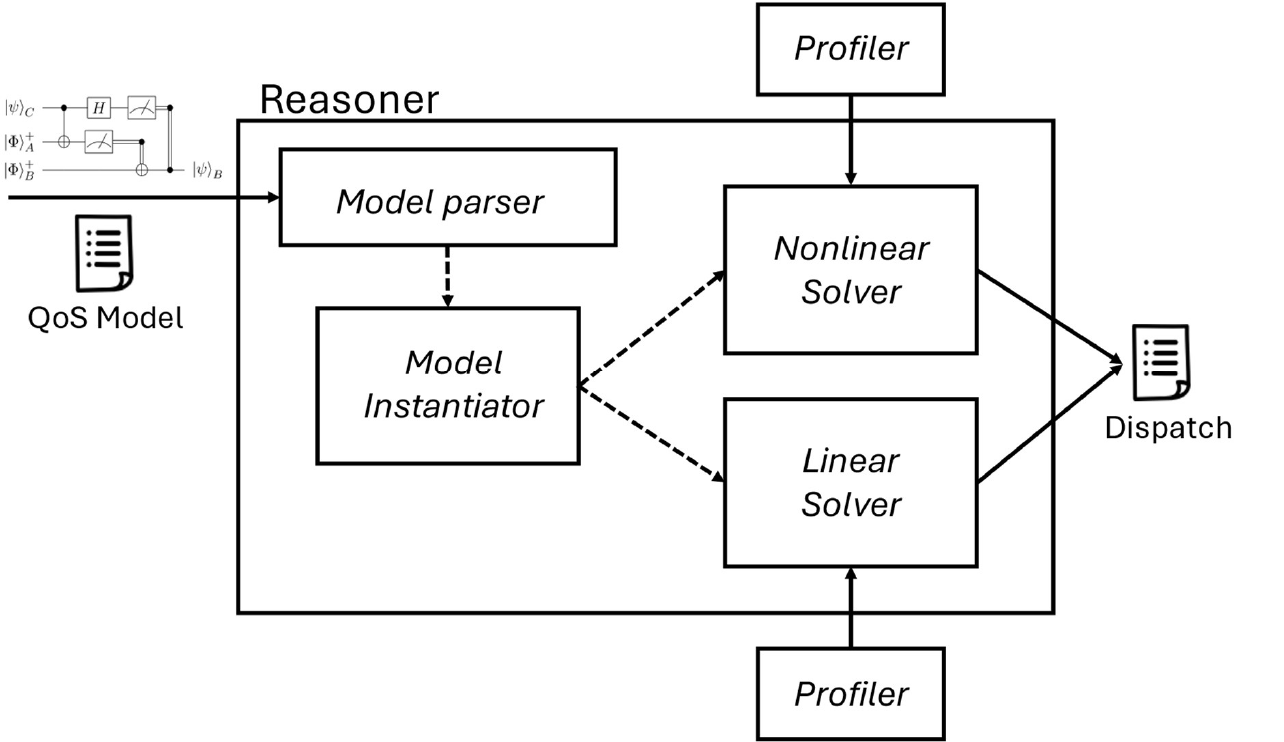}
{Architecture of Reasoner.\label{fig:reasonerarch}}

\paragraph{Model Parser and Model Instantiator}

Upon receiving as input a QoS model, a quantum circuit, and the total number of shots, the Reasoner first invokes its Model Parser module. This component interprets the \qoslanguageacr constraints and objective expressions, checks their consistency, and translates them into an internal representation suitable for the chosen solver backend.

Depending on the user-selected solver, the Model Instantiator enriches or transforms \qoslanguageacr expressions to better fit the solver's capabilities. \rev{For the MILP solver, cost and execution-time coefficients are obtained from one-shot Profiler estimates and multiplied by the corresponding integer shot variable, while waiting time is activated by the binary backend-use variable. The shot-dependent reliability quantity exposed through the QQSpec field \texttt{fidelity} is handled separately as a precomputed backend-level coefficient, as detailed below.} For the nonlinear solver, the Model Instantiator preserves the full form of the expressions, allowing for dynamic, instance-specific evaluation during optimization.

Additionally, the Model Instantiator expands any aggregate functions (such as \texttt{sum}, \texttt{max}, and \texttt{min}) over the set of candidate backends for both solver types. In the case of the linear model, this expansion involves introducing suitable auxiliary variables and additional linear constraints to accurately represent aggregates within the solver's formulation. The Model Instantiator also ensures that all implicit constraints, such as non-negativity and total shot allocation, are enforced in the final problem passed to the solver.

\new{To make the optimisation problem explicit, the Reasoner represents each dispatch through shot-allocation variables. 
Given a set of candidate backends \(B\) and a total shot budget \(S\), we define:}
\[
    x_b \in \mathbb{Z}_{\geq 0}
    \quad \text{as the number of shots assigned to backend } b \in B,
\]
and
\[
    y_b \in \{0,1\}
    \quad \text{as a binary variable indicating if backend } b \text{ is used.}
\]

\rev{Independently of the user-specified QoS model, the released MILP implementation enforces}
\begin{revision}
\[
\begin{aligned}
    \sum_{b \in B} x_b &= S,
    && \text{all requested shots are assigned,} \\
    0 \leq x_b &\leq S y_b,
    && \forall b \in B.
\end{aligned}
\]
\end{revision}

\rev{Together with the integer and binary domains above and \(S>0\), these constraints imply that at least one backend is used. No additional continuous relaxation or rounding step is introduced. A strictly positive lower bound on \(x_b\) for a used backend is added only when requested by QQSpec, for example \texttt{min(shots) >= 300}, which is preprocessed as \(x_b \geq 300 y_b\) for every backend.}

\rev{For the MILP used in the experiments, let \(c_b\) and \(t_b\) be the one-shot cost and execution-time coefficients returned by the Profiler, let \(w_b\) be the backend waiting-time estimate, and let \(\bar{R}_b=R_{\mathrm{syn},b}(S)\) be the synthetic reliability value precomputed at the total requested shot budget. When QQSpec specifies a minimum reliability threshold \(\tau\), the current implementation pre-filters candidate backends whose precomputed \(\bar{R}_b\) is below \(\tau\). For Scenario~1, where \(\tau=0.9\), let}
\begin{revision}
\[
    B_{0.9} =
    \left\{
        b \in B \mid \bar{R}_b \geq 0.9
    \right\}
\]
\end{revision}
\rev{denote the resulting eligible backend set. The Scenario~1 MILP surrogate is therefore constructed over \(B_{0.9}\) and includes}
\begin{revision}
\[
\begin{aligned}
    \sum_{b \in B_{0.9}} c_b x_b &\leq 800, \\
    x_b &\geq 300 y_b,
    && \forall b \in B_{0.9}, \\
    \max_{b \in B_{0.9}}
    (t_b x_b+w_b y_b)
    &\leq 2000.
\end{aligned}
\]
\end{revision}

\rev{The \texttt{max} operator is expanded by the implementation using an auxiliary variable and linear linking constraints. More generally, aggregate QQSpec operators are translated into solver-compatible expressions by the Model Instantiator. In the current MILP implementation, the QQSpec \texttt{fidelity} field is therefore not represented as a shot-dependent decision function inside the optimisation problem; instead, the precomputed backend-level reliability value is used to determine backend eligibility before the MILP is constructed.}

\rev{Similarly, with equal weights, the Scenario~1 objective is encoded using an auxiliary variable \(m\) for the maximum completion time:}
\begin{revision}
\[
\begin{aligned}
    \min_{x,y,m} \quad
        & 0.5\,m
        + 0.5 \sum_{b \in B_{0.9}} c_b x_b \\
    \text{s.t.} \quad
        & t_b x_b + w_b y_b \leq m,
        && \forall b \in B_{0.9}.
\end{aligned}
\]
\end{revision}

\rev{The difference between the two solvers lies in how the Profiler quantities are represented. The MILP solver uses linear coefficients for cost and execution time, a binary-activated waiting-time term, and a precomputed backend-level value for the QQSpec \texttt{fidelity} field. The Nonlinear Solver instead evaluates the original Profiler functions at each candidate shot allocation during the annealing search.}

\paragraph{Linear Solver (MILP-based)}

\rev{When instantiated, the Linear Solver module addresses the resource-allocation problem by formulating it as a mixed-integer linear program (MILP), rather than as a continuous linear program. For each backend, the model introduces an integer variable representing the number of assigned shots and a binary variable indicating whether that backend is used. These variables are linked through system-level constraints and are solved directly as integer and binary decision variables; the implementation does not rely on a continuous LP relaxation followed by a rounding procedure.}

In this setting, QoS metrics such as cost and execution time are represented through linear expressions of the allocation variables whenever possible. For each backend, the cost contribution is expressed using a backend-specific coefficient derived from the Profiler and the number of assigned shots, while analogous linear formulations are used for execution time. Waiting time is associated with backend activation and is therefore linked to the corresponding binary backend-use variable.

\rev{Metrics whose original profiling functions are nonlinear cannot in general be represented exactly within the MILP formulation. In particular, the shot-dependent reliability metric is represented through an approximation compatible with the linear optimisation model. In the current implementation, a backend-specific reliability value is computed by the Profiler before optimisation and subsequently used as a fixed backend-level quantity for filtering and constraint instantiation. Reliability requirements in the MILP therefore do not depend on the final shot allocation through the original nonlinear function. This preserves the efficiency and linear structure of the optimisation problem, but may cause the feasibility region seen by the solver to differ from that defined by the original Profiler.}

The resulting MILP incorporates all user-specified constraints and objectives, together with system-level implicit constraints such as conservation of the total number of shots and the coupling between shot-allocation and backend-use variables. The problem is solved using a mixed-integer optimisation solver, yielding an optimal solution, when one exists and optimality is certified by the solver, for the linearised mixed-integer formulation.

However, it is important to note that the solution produced by the Linear Solver is optimal only for the linear approximation of the original QoS model. The linearisation process introduces an inherent approximation, since it simplifies potentially nonlinear \qoslanguageacr expressions and backend profiling functions into relationships that can be represented within the MILP. As a result, the dispatch obtained from the optimisation model may not automatically satisfy the original, potentially nonlinear constraints and objectives specified by the user, nor fully capture the behaviour predicted by the original Profiler.

\rev{To prevent approximation errors from being exposed as valid QoS-aware dispatches, the Linear Solver includes a post-solve feasibility validation step. Once the MILP has produced a candidate shot allocation, the solver re-evaluates that fixed allocation using the original Profiler functions and checks the resulting QoS values against the original QQSpec constraints. A candidate that is feasible for the linearised model but violates the original profiling model is therefore rejected rather than being returned as a valid dispatch. We distinguish these cases as \emph{solver-feasible} but not \emph{QoS-feasible}.}

\rev{This check does not require execution of the quantum circuit or invocation of the Nonlinear Solver. Once the shot allocation has been fixed by the MILP, evaluating the original profiling functions for that candidate is a direct validation operation rather than a new optimisation problem, and therefore adds negligible overhead compared with nonlinear search.}

\rev{The current implementation performs validation but not automatic recovery. If a candidate fails the post-solve check, the Linear Solver reports that no validated dispatch was obtained. As future work, we plan to investigate lightweight repair heuristics that use the detected constraint violations and available QoS slack to adjust the candidate allocation before resorting to a more expensive nonlinear optimisation strategy.}

\rev{Accordingly, the optimality guarantee of the Linear Solver applies specifically to the MILP constructed from the linearised QoS representation. The solver does not claim global optimality with respect to the original nonlinear QoS formulation. Post-solve validation instead ensures that only allocations satisfying the original QQSpec constraints under the Profiler are exposed as valid dispatches.}

Overall, the Linear Solver is most suitable in scenarios where computational efficiency and deterministic optimisation are required and where the system behaviour can be adequately approximated through linear relationships. Its principal strength lies in efficient mixed-integer optimisation and guaranteed optimality for the linearised model, balanced against the approximation introduced when nonlinear QoS characteristics are represented within that model.

\paragraph{Nonlinear Solver (Dual Annealing)}

The Nonlinear Solver is designed to address resource allocation problems in which backend behaviour exhibits significant nonlinearity or context-dependent effects that cannot be reliably captured by linear models. This solver leverages a dual annealing optimisation process, which iteratively explores the space of possible dispatches using a combination of stochastic and local search strategies.

The workflow of the Nonlinear Solver proceeds in several stages. First, to provide an effective starting point for the annealing process, the solver conducts an initial search across all possible subsets of backends, evaluating candidate dispatches that divide the total number of shots equally among the selected backends. The most promising of these initial dispatches, according to the QoS objective, is chosen as the starting point for dual annealing.

\rev{The dual annealing phase consists of multiple stochastic annealing calls, each performing a configurable number of optimisation iterations. Both the number of calls and the iterations per call are user-configurable, allowing users to trade planning time against search effort. During each call, candidate dispatches are dynamically evaluated with the Profiler so that the original nonlinear synthetic QoS model is evaluated directly rather than through the MILP surrogate.}

For every candidate dispatch, all QoS constraints are immediately checked using the metrics returned by the Profiler. Only if the dispatch satisfies all constraints is the objective function computed. Across all iterations and annealing calls, the Nonlinear Solver tracks the feasible dispatch with the best (i.e., minimum or maximum, depending on the scenario) objective value obtained. The solution returned is the dispatch that achieves this best score.

While the Nonlinear Solver does not guarantee finding the global optimum, due to the stochastic nature of annealing and the size of the search space, it is able to explore a much broader and more realistic solution space than the Linear Solver. This enables it to capture the behaviour represented by the original profiling functions more accurately, particularly in the presence of nonlinear cost, time, and reliability effects. The trade-off for this accuracy is a longer computational time, governed by the user-selected annealing parameters.
\section{Experimental Evaluation}
\label{sec:experiments}

The objective of this experimental evaluation is to rigorously analyse the performance and flexibility of the Quantum Broker, with particular focus on how Quality of Service (QoS) modelling and dispatch policies influence resource allocation and solution quality. Specifically, we examine the impact of diverse QoS constraints and objective formulations on shot allocations and system behaviour, drawing on the scenarios introduced in Section~\ref{sec:qml}. In addition, we provide a comparative analysis of the Reasoner’s two solvers (linear and nonlinear), assessing their trade-offs in terms of computational efficiency, optimality, and fidelity to realistic execution conditions.

\subsection{Experimental Setting}
\label{sec:expsetting}

All experiments are conducted on a controlled, reproducible environment: a virtual machine equipped with an Intel(R) Xeon(R) Gold 5218 CPU @ 2.30GHz (8 cores, 32~GB RAM). Experimental code, configuration and results are publicly available\footnote{\url{https://github.com/di-unipi-socc/TheQuantumBroker-experiments}} for full reproducibility.


\new{To enable systematic and controlled evaluation of the Quantum Broker's dispatch and reasoning capabilities, we
decouple the experimental framework from direct access to live NISQ hardware while preserving realistic backend
heterogeneity. Specifically, we evaluate the Quantum Broker using noisy simulator backends corresponding to
real-device-inspired IBM fake backends\footnote{\new{IBM fake backends are simulator backends built from snapshots of real IBM Quantum systems. These
snapshots include device-specific information such as coupling maps, basis gates, qubit properties, and error
rates, and are commonly used for transpilation testing and noisy simulation.}}, namely \texttt{fake\_torino}, \texttt{fake\_kyiv},
\texttt{fake\_sherbrooke}, \texttt{fake\_fez}, and \texttt{fake\_marrakesh}. These fake backends are simulator models associated with snapshots of real quantum devices, including
hardware-specific information such as topology, basis gates, qubit properties, and noise-related calibration data.}

\newt{We intentionally avoid live-hardware evaluation because live queues, calibration drift, and provider availability would make solver comparisons non-repeatable. Our objective is to isolate the dispatch-planning behavior of the broker under controlled heterogeneous QoS conditions.}

\rev{In addition to this dispatch-planning evaluation, we perform a complementary end-to-end validation in which selected dispatches are actually executed on the device-specific noisy simulator backends, the measurement counts returned by the individual executions are merged, and the resulting output distribution is compared with an ideal noise-free reference.}

\new{In addition to the simulator noise model, each backend is associated with a performance profile used by the
Broker's Profiler to estimate QoS quantities during reasoning. These profiles encode backend-specific execution
and access characteristics, including base execution overhead, depth- and shot-dependent execution time,
queueing delay, price per unit time, and per-depth error. Thus, although the experiments are executed in a
controlled simulator environment, the simulated resources are intended to reflect the heterogeneity of real
quantum machines in both noise behaviour and operational performance.}

\new{Unlike the empirical profiler detailed in the previous section, which constructs profiles by actively benchmarking
live quantum hardware, the experimental profiler used here relies on fixed backend profiles associated with the
selected noisy simulator backends. This design gives us direct control over the experimental conditions while still
retaining real-device-inspired heterogeneity. The profile parameters include execution time, shot-dependent latency,
queueing delay, error rate, and usage cost, allowing the Reasoner to make dispatch decisions over backends that
differ in both noise and performance characteristics.}
This approach allows us to:
\begin{itemize}
\item \new{Systematically explore a broader variety of QoS regimes and stress scenarios than would be practical on
live hardware, while maintaining backend-specific noise and performance heterogeneity inspired by real
quantum machines.}
\item Precisely control and sweep relevant backend parameters to assess the Broker’s flexibility, robustness, and solution adaptability.
\item Facilitate direct comparison between different dispatch policies, solvers, and objective formulations. 
\item \new{Guarantee experiment reproducibility. Live hardware access is subject to stochastic queueing, calibration
drift, transient availability, and changing provider policies. In contrast, noisy simulator backends with fixed
QoS profiles enable repeatable experimentation: execution-time parameters, error rates, queueing delays, and
pricing models remain fixed across runs and are explicitly documented.}
\end{itemize}

The modular architecture of the Quantum Broker makes this substitution trivial: the Profiler component can be replaced at runtime, enabling us to swap between empirical hardware-driven profiling (for deployment) and fully controlled, synthetic profiling (for benchmarking and validation).

Table~\ref{tab:experiment-params} summarizes the key experimental settings. We leverage the MQT Bench framework to generate a diverse set of quantum circuits, instantiating seven representative quantum algorithms: Deutsch-Jozsa (dj), Quantum Approximate Optimisation Algorithm (qaoa), Quantum Fourier Transform (qft), Quantum Neural Network (qnn), RealAmplitudes random circuits (realamprnd), TwoLocal random circuits (twolocalrnd)~\cite{cross2019validating}, and Variational Quantum Eigensolver (vqe). For each algorithm family, circuits are generated at three sizes (6, 8, and 10 qubits) to examine scaling effects. The total number of shots per job is fixed at 4096.

The quantum backends consist of five Qiskit-based simulators (\texttt{fake\_torino}, \texttt{fake\_kyiv}, \texttt{fake\_sherbrooke}, \texttt{fake\_fez}, \texttt{fake\_marrakesh}), each assigned an independent, synthetic QPU profile. The Nonlinear Solver (dual annealing) is configured with 40 iterations per run and 10 independent runs for each optimization.

\begin{table*}[ht]
\centering
\caption{Experimental parameters used for evaluation.}
\label{tab:experiment-params}
\begin{tabular}{ll}
\hline
\textbf{Parameter}              & \textbf{Value / Setting}                                 \\
\hline
Total shots per job             & 4096                                                    \\
Backends             & fake\_torino, fake\_kyiv, fake\_sherbrooke, fake\_fez, fake\_marrakesh  \\
Nonlinear solver iterations     & 40                                              \\
Nonlinear solver runs           & 10                                                                      \\
Circuit algorithms              & dj, qaoa, qft, qnn, realamprandom, twolocalrandom, vqe  \\
Circuit sizes                   & 6, 8, 10 qubits                                         \\
\hline
\end{tabular}
\end{table*}

The adoption of noisy simulators and synthetic profiles is a deliberate experimental decision: synthetic profiling enables us to emulate a wide range of realistic and adversarial backend scenarios. For each backend, we specify, in a controlled and documented manner, static and dynamic parameters such as base execution time, time per shot per circuit depth, queue waiting time, unit price, and per-depth error rate. This maximises flexibility in experimental design and allows us to stress-test the Broker’s behaviour across edge-case and representative workload regimes.

\rev{The term ``simulator backend'' should therefore be interpreted carefully in this evaluation. We do not use
noise-free simulators as interchangeable computational resources. Instead, each backend emulates a specific
machine-like configuration and is paired with a QoS profile that captures its expected performance and reliability.
The noisy simulator accounts for device-level execution noise, while the profile supplies the operational quantities
needed by the Broker, such as time, cost, waiting time, and the synthetic reliability score defined below. This separation mirrors the Broker
architecture: execution behaviour is represented by the backend, whereas dispatch-time decision making is driven
by the Profiler's QoS abstraction.}

\rev{The backends' profiling values are summarised in Table~\ref{tab:backend-profiles}. These profiles are used by the synthetic profiler to compute expected execution time, cost, and the synthetic reliability score \(R_{\mathrm{syn}}\) for each backend, given a quantum circuit and shot allocation. The interpretation and use in the evaluation of the profiles are as follows:}

\begin{table*}[ht]
\centering
\caption{Profiles of the simulator backends used in the experiments.}
\label{tab:backend-profiles}
\begin{tabular}{lccccc}
\hline
\textbf{Backend} & \textbf{Base Time} & \textbf{Time/Shot/Depth} & \textbf{Waiting Time} & \textbf{Price/Time} & \textbf{Error/Depth} \\
\hline
fake\_torino      & 0.10      & 0.05        & 75          & 0.9     & 0.015  \\
fake\_kyiv        & 0.09      & 0.06        & 90          & 0.5     & 0.020  \\
fake\_sherbrooke  & 0.07      & 0.10        & 85          & 0.2     & 0.016  \\
fake\_fez         & 0.12      & 0.10        & 30          & 0.6     & 0.018  \\
fake\_marrakesh   & 0.04      & 0.08        & 85          & 0.5     & 0.010  \\
\hline
\end{tabular}
\end{table*}

\begin{itemize}
\item \textbf{Base Time}: The fixed execution overhead required to initialise and process a quantum job on the backend.
\item \textbf{Time/Shot/Depth}: The incremental execution time added for each shot and each unit of circuit depth, indicating how execution time grows with workload and circuit complexity.
\item \textbf{Waiting Time}: The expected delay before job execution can begin, representing queueing time for the backend.
\item \textbf{Price/Time}: The cost incurred per unit of execution time on the backend, reflecting the backend's usage pricing policy.
\item \textbf{Error/Depth}: \rev{The synthetic backend-error contribution per unit of circuit depth used in \(R_{\mathrm{syn}}\). It is a model parameter for the controlled evaluation, not an empirical distributional-fidelity measurement.}
\end{itemize}

\rev{These profiling parameters allow the evaluation framework to generate backend-specific estimates for execution time and cost and a controlled synthetic reliability score for each circuit and shot allocation. The reliability score is an experimental QoS proxy; it must not be interpreted as the empirical distributional fidelity \(F_{\mathrm{BC}}\) defined for the empirical Profiler.}

\paragraph{Profiling Methodology and Formulas}

For each backend, the synthetic profiler calculates the relevant metrics as follows:

\rev{For compactness, we denote the backend parameters reported in Table~\ref{tab:backend-profiles} by \(T_{\mathrm{base}}\) (Base Time), \(\gamma\) (Time/Shot/Depth), \(T_{\mathrm{wait}}\) (Waiting Time), \(\rho\) (Price/Time), and \(\epsilon\) (Error/Depth). These symbols are only mathematical notation for the corresponding profile fields.}

\begin{itemize}
    \item \textbf{Execution time:} \rev{For a circuit of depth \(d\) and a positive shot count \(s\), the synthetic evaluator uses}
    \[
    \rev{T_{\mathrm{exec}} = T_{\mathrm{base}} + \gamma d s.}
    \]
    \rev{The first term represents a fixed backend overhead, while the second increases with both circuit depth and shot count.}
    
    \item \textbf{Waiting time:} The waiting time $T_{\mathrm{wait}}$ is a fixed value per backend, set by the profile as \texttt{waiting\_time}.
    
    \item \textbf{Cost:} \rev{The synthetic evaluator computes the total job cost as}
    \[
    \rev{C = \rho\, T_{\mathrm{exec}}^{\beta},}
    \]
    \rev{where \(\rho\) is backend-specific and \(\beta=0.9\) in the experiments. The sublinear exponent is used to emulate effects such as bulk execution or amortised overhead; it is a controlled pricing abstraction rather than a universal provider pricing model.}
    
    \item \textbf{Synthetic reliability score.}
    \rev{The dispatch-planning experiments use a synthetic QoS reliability score that combines a depth-dependent backend-error term with a finite-shot uncertainty scale. This quantity is distinct from the empirical output-distribution fidelity evaluated in the end-to-end experiment described in Section~\ref{sec:end-to-end}. For a backend with per-depth error parameter \(\epsilon\geq 0\), a circuit of depth \(d\geq 0\), and a positive integer shot allocation \(s\in\mathbb{N}_{>0}\), we define}

    \[
    \rev{R_{\mathrm{syn}}
    =
    1-
    \sqrt{
    (\epsilon d)^2
    +
    \left(
    \frac{1}{2\sqrt{s}}
    \right)^2
    }.}
    \]

    \rev{The term \(\epsilon d\) is a synthetic depth-dependent degradation proxy, whereas \(1/(2\sqrt{s})\) is a simple finite-sampling uncertainty scale that decreases as the shot count increases. Their quadrature combination is a modelling choice for the controlled evaluation; it is not a physical error-composition law, the squared Bhattacharyya coefficient, or a quantum-state fidelity. The formula is evaluated only for used backends with \(s>0\). Consistently with the implementation, an unused backend with \(s=0\) is assigned zero QoS values rather than evaluating the expression, for which the sampling term would be undefined.}

    \rev{Accordingly, \(R_{\mathrm{syn}}\) should be interpreted as a synthetic QoS score whose larger values indicate a more favourable combination of backend degradation and sampling uncertainty. It is upper-bounded by \(1\), but it is not constrained to be non-negative: whenever the combined penalty exceeds one, the expression yields \(R_{\mathrm{syn}}<0\). Such a value does not represent a negative physical fidelity; it only indicates that the synthetic penalty has exceeded the unit reference level of the score. We therefore do not interpret \(R_{\mathrm{syn}}\) as a quantity restricted to \([0,1]\).}

    \rev{By contrast, the empirical Profiler described earlier measures output-distribution similarity with \(F_{\mathrm{BC}}\) and fits those measurements to obtain the dispatch-time prediction \(F_{\mathrm{dispatch}}\). The current QQSpec implementation exposes both profiler-specific quantities through the field named \texttt{fidelity}; in the controlled experiments and their scenario specifications, that field contains \(R_{\mathrm{syn}}\). Accordingly, labels such as ``fidelity'' in the experimental figures and QQSpec listings refer to the synthetic reliability score, not to \(F_{\mathrm{BC}}\). The end-to-end experiment in Section~\ref{sec:end-to-end}, instead, evaluates the empirical distribution produced after execution and shot aggregation and therefore reports a distinct metric, \(F_{\mathrm{E2E}}\).}

\end{itemize}

\rev{This synthetic profiling approach enables us to simulate a heterogeneous quantum resource pool with diverse performance and economic characteristics. The two solvers leverage these profiles differently: the Nonlinear Solver employs the profiles dynamically; at each dual annealing iteration, it queries the synthetic profiler to obtain execution time, cost, and \(R_{\mathrm{syn}}\) for the current dispatch configuration and circuit, thus capturing potential nonlinear or context-dependent effects that arise in practice.}
\rev{In contrast, the Linear Solver derives linear cost and execution-time coefficients from the profile parameters. Reliability is handled separately: for each backend, a synthetic reliability value \(R_{\mathrm{syn},b}(S)\) is precomputed at the total requested shot budget \(S\) and used as a fixed backend-level quantity for filtering and constraint instantiation. The resulting dispatch is subsequently re-evaluated using the original shot-dependent Profiler functions during post-solve validation. Although the Linear Solver uses linear approximations within its optimisation model, its final dispatch is evaluated using the same synthetic profiling formulas used to evaluate the Nonlinear Solver's dispatches, ensuring a fair and consistent comparison of cost, execution time, waiting time, and \(R_{\mathrm{syn}}\).}

\paragraph{\rev{Alignment with real hardware and scope of the synthetic model.}}

\rev{The synthetic profiling models defined above are intended as controlled abstractions of qualitative trends observed in cloud-accessible quantum computing, rather than as complete physical or economic models of specific devices. Their purpose is to provide repeatable backend heterogeneity for evaluating the Broker's reasoning and dispatch behaviour under well-defined QoS conditions.}

\rev{The execution-time model represents a fixed backend overhead together with a workload-dependent component that increases with circuit depth and shot count. This abstraction is sufficient for testing whether the Reasoner reacts correctly to differences in backend speed and workload size. Real execution time may additionally depend on provider-specific compilation pipelines, batching policies, hardware scheduling, control-system overheads, queue dynamics, and backend availability.}

\rev{The cost model introduces a backend-specific price coefficient and sublinear scaling through \(\beta<1\), thereby emulating effects such as bulk execution or amortised overhead. It is not intended as a universal pricing law: production pricing may instead depend on per-shot charges, reserved access plans, task-submission overheads, credits, or other provider-specific policies.}

\rev{The synthetic reliability score captures only the two controlled effects encoded in its definition: increasing degradation with circuit depth and decreasing finite-shot uncertainty with increasing shot count. Because \(R_{\mathrm{syn}}\) is a synthetic score rather than a physical fidelity, values below zero are permitted by the model and should be interpreted only as stronger synthetic degradation, not as negative distributional fidelity. Real-hardware output quality can deviate substantially from this proxy because it also depends on gate composition, qubit mapping, connectivity, calibration drift, readout error, crosstalk, temporal noise fluctuations, transpilation choices, and error mitigation.}

\rev{The synthetic profiles should therefore be interpreted as controlled experimental profiles, not as a validation that the same numerical QoS values would be obtained on live hardware. In a production deployment, these assumptions can be corrected by replacing the synthetic profile with an empirical profile obtained from real backend observations. This is already supported by the architecture of the Quantum Broker: the Profiler is modular, and the Reasoner only requires backend-level estimates of cost, execution time, waiting time, and fidelity.}

\rev{A practical calibration procedure would consist of four steps. First, benchmark circuits with varying depths, widths, gate compositions, and shot counts are executed on each target backend. Second, the observed execution time, queue delay, cost, and output-distribution fidelity are recorded. Third, the parameters of the profiling model are fitted from these observations, for example by estimating \(T_{\mathrm{base}}\), \(\gamma\), \(\rho\), \(\beta\), and \(\epsilon\) through regression. Fourth, the fitted profile is validated on held-out circuits not used during calibration. If systematic residual errors are observed, the analytical formulas can be replaced with richer models, such as gate-aware fidelity predictors, backend-specific timing models, interpolation tables, or learned regressors.}

\rev{This separation between the Broker and the profiling model is important. The contribution of the Quantum Broker is not tied to one specific synthetic formula; rather, the formulas provide a reproducible experimental substrate for evaluating QoS-aware shot allocation. More accurate real-hardware profiles can be incorporated without changing the QQSpec language, the Reasoner interface, or the Executor.}


\paragraph{Baselines.}
\rev{To provide a meaningful comparison for the Reasoner's solvers, we consider seven baseline dispatch strategies. Five baselines assign all 4096 shots exclusively to a single backend, one for each available backend: \texttt{all\_fez}, \texttt{all\_kyiv}, \texttt{all\_marrakesh}, \texttt{all\_sherbrooke}, and \texttt{all\_torino}. The \texttt{fair} baseline distributes the shots evenly among all five backends. The \texttt{qos-pareto} baseline provides a simple QoS-aware multi-backend heuristic. Each backend is evaluated under the same fair-share reference allocation, after which backends that violate backend-local hard requirements or are Pareto-dominated in the QoS dimensions relevant to the scenario are discarded. Backend-specific requirements expressed in QQSpec take precedence over this filtering: a backend that is explicitly required by the specification is retained rather than removed solely because it is Pareto-dominated. The total shot budget is then divided uniformly among the surviving backends.}

\rev{These baselines cover complementary dispatch strategies under the same candidate pool. The \texttt{all\_*} baselines enumerate every possible single-backend allocation, and therefore span the decision space of methods that ultimately select one backend, or one backend--circuit/compiler configuration, for the entire workload. The \texttt{fair} baseline represents multi-backend distribution without considering QoS characteristics, whereas \texttt{qos-pareto} introduces lightweight QoS awareness through eligibility and dominance filtering without performing allocation optimisation. For all baselines, including \texttt{qos-pareto}, the final dispatch is evaluated against the complete original QQSpec model, including aggregate constraints and backend-specific allocation requirements. If any requirement is violated, the dispatch is reported as infeasible; \texttt{qos-pareto} does not perform further search, repair, or objective optimisation.}

\rev{All baselines are evaluated using the same circuits, backend profiles, QoS metrics, and objective functions used for the Quantum Broker solvers. This ensures that differences in performance are attributable to the dispatch strategy itself, rather than to differences in provider state, calibration drift, implementation details, or benchmark setup.}

\new{In this initial work, we explicitly decided not to include direct quantitative comparisons with existing systems. A direct quantitative comparison with existing state-of-the-art quantum execution managers is not straightforward, because the Quantum Broker addresses a different optimisation problem from most prior systems. Existing approaches typically perform \emph{single-backend selection}: given a circuit, they identify one suitable QPU, compiler configuration, or QPU-circuit pair, and then execute the entire shot budget on that selected backend. This is the case for systems such as the Quantum API Gateway, the NISQ Analyzer family, and related adaptive scheduling approaches, which optimise backend or configuration choice but do not formulate shot allocation across multiple heterogeneous QPUs as the primary decision variable.}

\new{In contrast, the Quantum Broker solves a \emph{shot-wise dispatch optimisation} problem. Its output is not a single selected backend, but a dispatch plan assigning portions of the same circuit's shot budget to multiple backends subject to declarative QoS constraints and objectives. Consequently, existing systems are not directly comparable without modifying their problem formulation, extending their implementation to support shot-wise execution, and redefining their objective functions and constraint semantics. Such modifications would effectively turn them into variants of our proposed approach, making the comparison methodologically ambiguous.}

\rev{For this reason, we compare against controlled baselines spanning both single- and multi-backend dispatch. The \texttt{all\_*} baselines cover the complete single-backend decision space, while \texttt{fair} and \texttt{qos-pareto} provide multi-backend reference policies without solving an optimisation problem. This setup separates the effects of the dispatch policy from those of explicit constraint-based optimisation.}

\rev{This baseline design isolates the contribution of the Quantum Broker's QoS-aware shot allocation without confounding the comparison with differences in implementation availability, provider access, calibration state, compiler stack, or benchmark configuration. The evaluation therefore compares optimisation-based shot-wise reasoning, under identical backend profiles and QoS specifications, with exhaustive single-backend choices and the two multi-backend baseline policies.}

\new{In the following evaluation, we deliberately use a small number of simple benchmark scenarios to avoid confounding the evaluation with application-specific policy design. The goal of the experiments is to assess whether the Quantum Broker can correctly translate declarative QoS requirements into feasible and high-quality shot allocations, and to compare the behaviour of the linear and nonlinear solvers under identical conditions. Simple scenarios make constraint violations, objective-value differences, and solver failure modes easier to interpret.}

\new{This does not restrict the framework to simple use cases. QQSpec allows users to define more elaborate policies by combining the same variables and operators used in the benchmark scenarios. The scenarios should therefore be read as controlled experimental probes of the Broker's capabilities, rather than as the limit of the policies that the system can express.}

\rev{Finally, the absence of direct quantitative comparison with prior orchestration frameworks follows from the mismatch in problem formulation. Existing systems primarily address backend selection, whereas Quantum Broker addresses constrained shot-wise allocation across multiple backends. Since every single-backend selector must choose one backend from the candidate set, the exhaustive \texttt{all\_*} baselines represent the complete set of possible single-backend dispatches under the experimental conditions. The multi-backend baselines provide complementary reference points without invoking the Broker's MILP formulation or nonlinear search.}

\subsection{Scenario 1 Analysis}

\Figure[!ht][width=\textwidth]{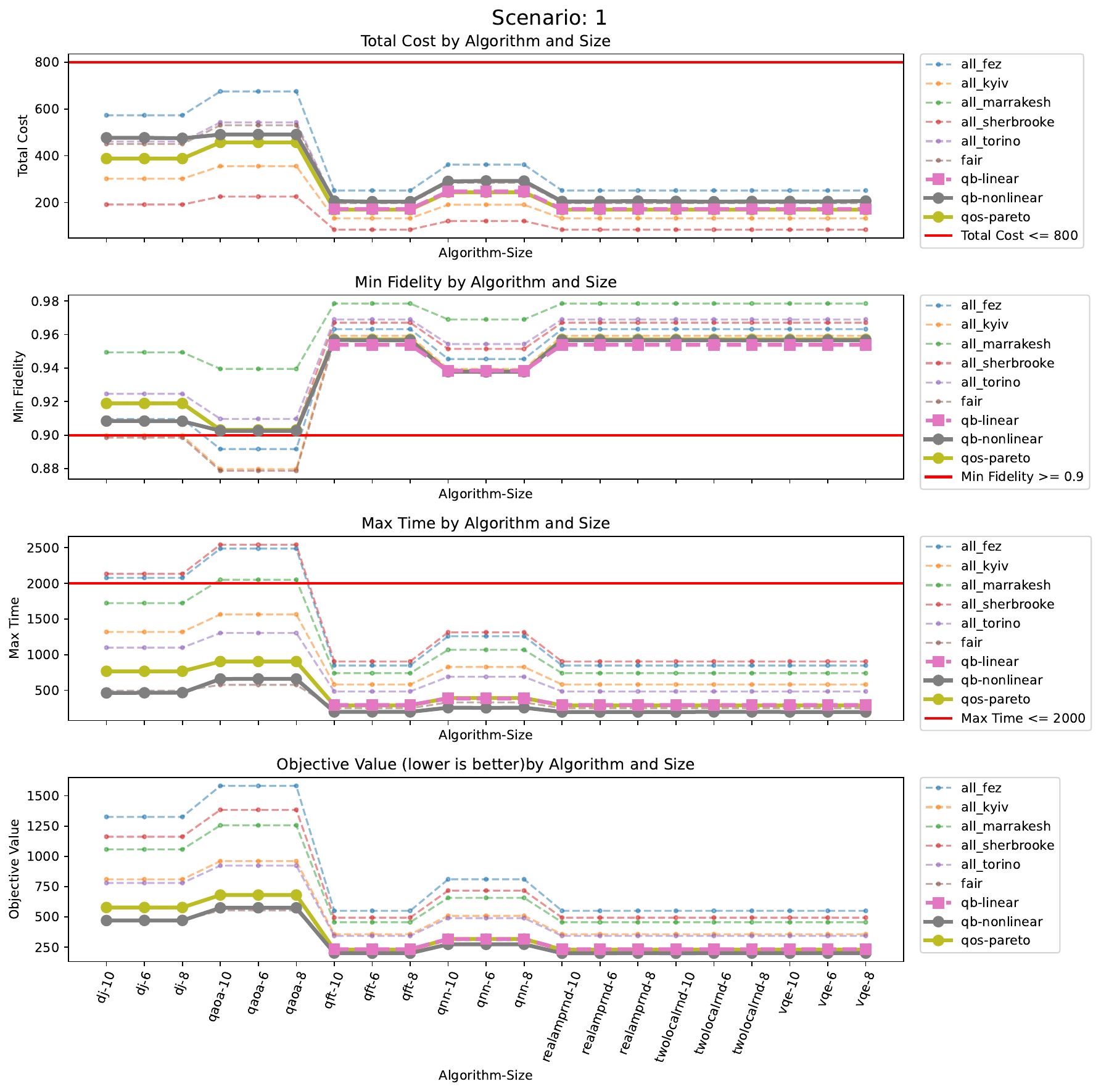}
{\rev{Results for Scenario 1. The panels report total cost, minimum fidelity, maximum completion time, and objective value for the single-backend baselines, \texttt{fair}, \texttt{qos-pareto}, and the two Reasoner solvers. Horizontal red lines mark the QoS thresholds.} \label{fig:scenario1}}

\rev{Figure~\ref{fig:scenario1} presents the results for Scenario 1 (Listing~\ref{model:scenario1}), displaying, for each algorithm and circuit size, total cost, minimum fidelity, maximum execution time, and the objective value. The comparison includes the five single-backend baselines, \texttt{fair}, \texttt{qos-pareto}, \texttt{qb-linear}, and \texttt{qb-nonlinear}. Horizontal red lines identify the cost, fidelity, and completion-time thresholds.}

\rev{Across the evaluated instances, \texttt{qos-pareto} remains below the cost and time limits and above the minimum-fidelity threshold, including the \texttt{dj} and \texttt{qaoa} workloads for which several single-backend allocations violate at least one QoS requirement. Its completion times and objective values are also generally lower than those of the \texttt{fair} and single-backend baselines.}

Each point on the x-axis corresponds to a particular algorithm and circuit size, as generated by the MQT Bench framework\cite{quetschlich2023mqt}. For every instance, the corresponding dispatch is evaluated using the profiling methodology detailed in Section~\ref{sec:expsetting}.

\rev{Figure~\ref{fig:scenario1} also shows that \texttt{qb-linear} fails to find feasible solutions for the \texttt{dj} and \texttt{qaoa} circuits. For these instances, the linear surrogate makes the approximated cost, time, and reliability constraints infeasible in the MILP model. Both \texttt{qos-pareto} and \texttt{qb-nonlinear} instead produce feasible dispatches under the original Profiler evaluation.}

\rev{For the objective values in the bottom panel, \texttt{qb-nonlinear} generally attains lower values than \texttt{qos-pareto} on the most constrained \texttt{dj} and \texttt{qaoa} instances. For the remaining workloads, their values are closer, consistent with the larger feasible region.}

\rev{Scenario 1 shows that QoS-based filtering is sufficient for several workloads, while tighter constraints and interacting QoS quantities can still benefit from optimisation over the complete shot allocation.}

\subsection{Scenario 2 Analysis}

\Figure[!ht][width=\textwidth]{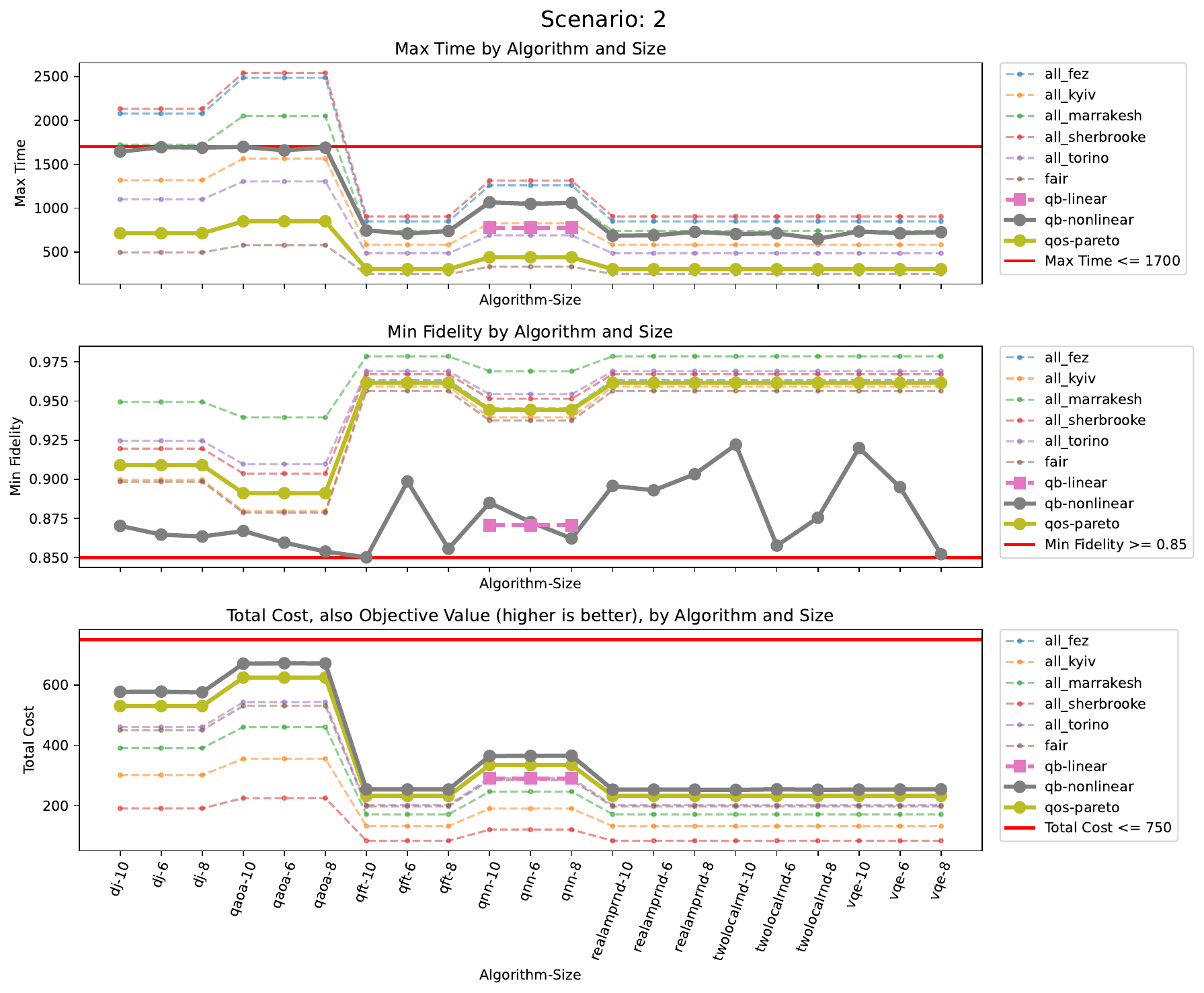}
{\rev{Results for Scenario 2. The panels report maximum completion time, minimum fidelity, and total cost (the objective to maximise) for the single-backend baselines, \texttt{fair}, \texttt{qos-pareto}, and the two Reasoner solvers. Horizontal red lines mark the QoS thresholds.} \label{fig:scenario2}}

\rev{Figure~\ref{fig:scenario2} presents the results for Scenario 2 (Listing~\ref{model:scenario2}), where the QoS bounds are less restrictive than in Scenario 1. The three panels report maximum execution time, minimum fidelity, and total cost, which is also the budget-utilisation objective to maximise.}

\rev{The single-backend baselines benefit from the broader constraint margins, although some allocations still exceed the maximum-time limit. \texttt{qos-pareto} remains within the displayed time and fidelity limits across the evaluated instances.}

\rev{The contrast with \texttt{qb-linear} is particularly visible in the fidelity panel. Although the MILP solver returns a candidate for every instance, its linearised reliability representation leads to allocations that fail the original fidelity requirement when re-evaluated with the Profiler. The reported \texttt{qos-pareto} allocations remain feasible under the same evaluation.}

\rev{For the objective in the bottom panel, \texttt{qos-pareto} generally achieves higher budget utilisation than \texttt{fair} and the single-backend baselines, while \texttt{qb-nonlinear} obtains the highest or near-highest feasible objective values across the workloads.}

\rev{Scenario 2 shows that \texttt{qb-nonlinear} can improve budget utilisation beyond QoS-based filtering while maintaining feasibility under the original Profiler model.}

\rev{This behaviour motivates an explicit post-solve feasibility validation step. A dispatch returned by a solver is initially feasible only with respect to that solver's internal optimisation model. Before being considered QoS-feasible, the candidate dispatch is therefore re-evaluated using the original Profiler functions and checked against the QQSpec constraints. In Scenario~2, the Linear Solver returns candidates that are feasible for its linearised model, but the post-solve evaluation detects that the original reliability constraint is violated. These candidates are consequently classified as solver-feasible but not QoS-feasible and must not be exposed as validated dispatches. Importantly, this validation does not require executing the circuit or invoking the Nonlinear Solver; once the shot allocation is known, the original profiling functions can be evaluated directly on that candidate dispatch.}

\subsection{Scenario 3 Analysis}

\Figure[!ht][width=\textwidth]{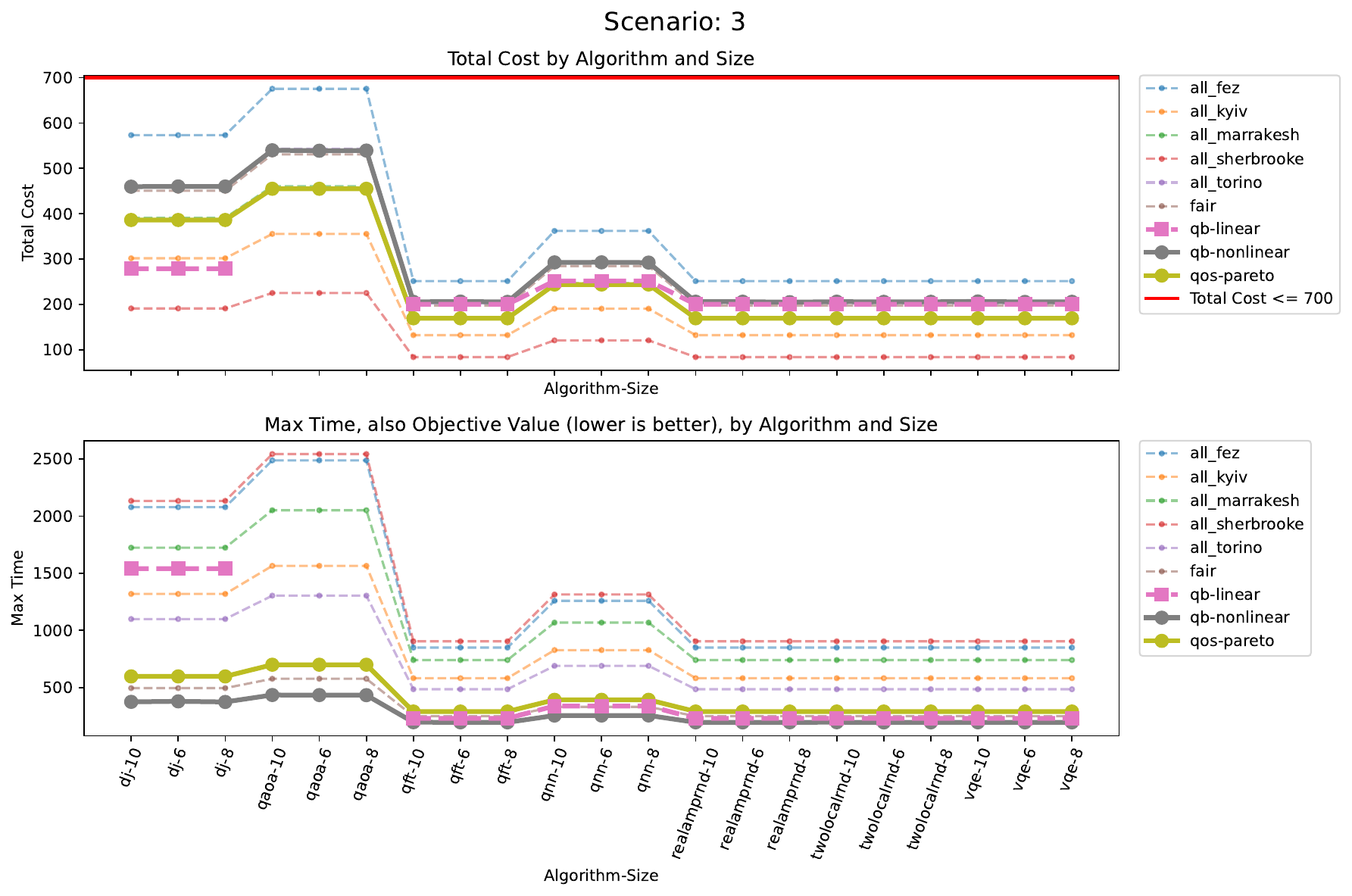}
{\rev{Results for Scenario 3. The panels report total cost and maximum completion time (the objective to minimise) for the single-backend baselines, \texttt{fair}, \texttt{qos-pareto}, and the two Reasoner solvers. The red line marks the total-cost limit.} \label{fig:scenario3}}

\rev{Figure~\ref{fig:scenario3} presents the evaluation for Scenario 3 (Listing~\ref{model:scenario3}), where the objective is to minimise the overall completion time subject to a total-cost limit of 700 and the requirement that at least 700 shots are assigned to \texttt{fake\_torino}. The two panels report total cost and maximum completion time for each circuit instance.}

\rev{The targeted backend constraint distinguishes this scenario from the previous ones. Among the single-backend baselines, only \texttt{all\_torino} can satisfy the mandatory \texttt{fake\_torino} assignment, while \texttt{fair} assigns enough shots to that backend through its uniform split. The \texttt{qos-pareto} baseline also satisfies the mandatory assignment, and all of its reported dispatches remain below the cost ceiling.}

\rev{\texttt{qos-pareto} substantially improves the maximum completion time with respect to the single-backend strategies and the \texttt{fair} allocation. The Reasoner outputs generally achieve lower completion times still, because they optimise the number of shots assigned to each selected backend instead of using a uniform split.}

\rev{As in Scenario 1, \texttt{qb-linear} does not return feasible candidates for the \texttt{qaoa} instances under its linearised model. Where solver outputs are available, \texttt{qb-nonlinear} achieves the lowest or among the lowest maximum completion times.}

\rev{Scenario 3 shows that QoS-based filtering can reduce completion time substantially, while explicit shot-allocation optimisation provides further gains when backend-specific constraints and completion-time trade-offs interact.}

\subsection{End-to-End Dispatch Execution Validation}
\label{sec:end-to-end}

\rev{The experiments above assess whether the dispatches generated by the Reasoner satisfy the QoS requirements according to the profiles available at dispatch time. To complement this planning-level analysis, we perform an additional end-to-end experiment that executes selected multi-backend dispatches, collects the measurement results returned by each backend, merges them, and evaluates the quality of the final output distribution. This experiment directly addresses whether a dispatch that has been selected by the Quantum Broker can be carried through the complete execution and aggregation workflow while preserving the statistical output of the original quantum computation.}

\rev{In this end-to-end experiment, we focus on output-distribution quality rather than on execution time or monetary cost. The wall-clock time required to simulate a circuit on a classical machine is not representative of the execution time of the corresponding QPU, and simulated backends do not reproduce provider-side queueing, availability, or billing mechanisms. Consequently, using simulator runtime or simulator resource consumption as empirical estimates of these QoS quantities would be misleading. Execution time, waiting time, and cost therefore remain evaluated through the controlled backend profiles introduced in Section~\ref{sec:expsetting}. In contrast, the noisy simulator directly produces backend-specific measurement outcomes, making output-distribution quality the QoS dimension that can be meaningfully assessed at execution level in the present experimental setting.}

\rev{For this validation, we focus on the common subset of workloads for which the Nonlinear Solver produces a feasible dispatch in all three scenarios, namely the 6-, 8-, and 10-qubit Deutsch-Jozsa and QFT instances. This results in 18 end-to-end executions, corresponding to six common workloads evaluated under each of the three scenarios. Using the same workload set across all scenarios enables a direct comparison of end-to-end execution quality without confounding differences between scenarios with changes in workload composition.}

\rev{For each selected dispatch, the circuit is independently transpiled for every backend receiving a non-zero shot allocation. The transpilation therefore respects the topology, basis gates, and device properties of the corresponding IBM fake backend. We then instantiate a noisy Qiskit Aer simulator from each fake backend and execute exactly the number of shots assigned to that backend by the Reasoner. If \(s_b\) is the number of shots assigned to backend \(b\), and \(c_b(x)\) is the number of occurrences of output bit string \(x\) returned by that backend, the Broker reconstructs the final result by summing the partial counts:}

\begin{revision}
\[
c_{\mathrm{merged}}(x)
=
\sum_{b \in B_{\mathrm{used}}} c_b(x),
\]
\end{revision}

\rev{where \(B_{\mathrm{used}}\) is the set of backends receiving at least one shot. Since every dispatch preserves the total shot budget,}

\begin{revision}
\[
\sum_x c_{\mathrm{merged}}(x)
=
\sum_{b \in B_{\mathrm{used}}} s_b
=
4096.
\]
\end{revision}

\rev{The application-visible empirical distribution returned after aggregation is therefore}

\begin{revision}
\[
p_{\mathrm{merged}}(x)
=
\frac{c_{\mathrm{merged}}(x)}{4096}.
\]
\end{revision}

\rev{To obtain an independent reference, we execute the same logical circuit with a noise-free simulator and derive its ideal output distribution \(p_{\mathrm{ideal}}\). We then quantify end-to-end output quality using the squared Bhattacharyya coefficient}

\begin{revision}
\[
F_{\mathrm{E2E}}
=
\left(
\sum_{x \in \mathcal{X}}
\sqrt{
p_{\mathrm{merged}}(x)
p_{\mathrm{ideal}}(x)
}
\right)^2,
\]
\end{revision}

\rev{where \(F_{\mathrm{E2E}}\in[0,1]\) and values close to one indicate strong agreement between the output distribution produced by the complete shot-wise execution and the ideal reference. Importantly, \(F_{\mathrm{E2E}}\) is not the same quantity as \(R_{\mathrm{syn}}\). The latter is a dispatch-time QoS estimate used by the Reasoner, whereas \(F_{\mathrm{E2E}}\) is measured only after the circuit has been transpiled, executed under backend-specific noise, and the returned shots have been merged. Consequently, this experiment provides an execution-level validation that does not simply re-evaluate the selected dispatch using the same model employed during optimisation.}

\begin{table}[htbp]
\centering
\begin{tabular}{lccc}
\toprule
\textbf{Workload} &
\textbf{Scenario 1} &
\textbf{Scenario 2} &
\textbf{Scenario 3} \\
\midrule
dj-6   & 0.913 & 0.926 & 0.901 \\
dj-8   & 0.827 & 0.870 & 0.818 \\
dj-10  & 0.796 & 0.840 & 0.756 \\
qft-6  & 0.998 & 0.995 & 0.997 \\
qft-8  & 0.985 & 0.983 & 0.984 \\
qft-10 & 0.913 & 0.919 & 0.911 \\
\midrule
\textbf{Mean} & \textbf{0.905} & \textbf{0.922} & \textbf{0.895} \\
\bottomrule
\end{tabular}
\caption{\rev{End-to-end squared Bhattacharyya coefficient \(F_{\mathrm{E2E}}\) between the merged 4096-shot output distribution and the ideal reference distribution. The table reports the common subset of workloads for which the Nonlinear Solver produces a feasible dispatch in all three scenarios, allowing a direct comparison across identical circuit instances.}}
\label{tab:e2e-fidelity}
\end{table}

\rev{Table~\ref{tab:e2e-fidelity} reports the resulting aggregate-distribution quality for the common feasible workload subset. Across the 18 end-to-end executions, the mean \(F_{\mathrm{E2E}}\) is \(0.907\), with values ranging from \(0.756\) to \(0.998\). The mean values are \(0.905\), \(0.922\), and \(0.895\) for Scenarios 1, 2, and 3, respectively. Overall, the results show that the selected dispatches can be carried through the complete execution and aggregation workflow while preserving substantial agreement with the corresponding ideal output distributions.}

\rev{The results also expose an important distinction between the quality of an individual partial execution and that of the final result returned by the Broker. This effect is particularly visible for QFT, whose output probability mass is spread over many bit strings. In Scenario 2, for example, the 10-qubit QFT dispatch assigns only 76 shots to one of the participating backends. The minimum distributional similarity measured independently over the individual backend executions is approximately \(0.073\), whereas the distribution reconstructed after merging all 4096 shots achieves \(F_{\mathrm{E2E}}\approx0.919\). The low backend-local value is therefore not representative of the application-visible result: a backend receiving only a small subset of the shots cannot by itself reconstruct a broad output distribution, even in the absence of substantial device noise.}

\rev{This observation is directly relevant to the shot-wise execution model of the Quantum Broker. Individual backends do not produce independent final answers; instead, they contribute samples that are subsequently combined into a single output distribution. Backend-level quantities remain useful during resource selection and QoS reasoning, but the merged distribution provides the appropriate execution-level measure of the result delivered to the user. The experiment therefore confirms that evaluating only isolated partial executions can underestimate the quality achieved after shot aggregation.}

\rev{Finally, this end-to-end experiment should not be interpreted as a substitute for live-QPU validation. The measurement outcomes are obtained from device-specific noisy simulators, while queue delay, execution time, cost, and other operational QoS quantities remain determined by the controlled profiles described in Section~\ref{sec:expsetting}. The experiment validates the dispatch--execution--aggregation mechanism and the resulting output quality under reproducible backend-specific noise; empirically measured operational profiles and execution on live quantum processors remain complementary validation steps.}

\subsection{Linear Solver vs Nonlinear Solver}
To assess the practical effectiveness of the Reasoner, we compare its two solvers across a diverse set of quantum workload scenarios. This comparison aims to illustrate the trade-offs between computational speed, solution optimality, and robustness to model approximations. By analysing how each solver performs under identical conditions, we seek to clarify in which contexts the computational efficiency of the MILP formulation outweighs its limitations, and when the increased realism and adaptability of dual annealing justifies its greater computational cost.
\rev{The Linear Solver provides an exact optimisation reference within the
scope of its own formulation. Since the instantiated problem is a
mixed-integer linear program (MILP), the solution returned by the
mixed-integer optimisation solver is globally optimal with respect to
the linearised QoS formulation when optimality is certified by the
solver. This guarantee applies only to the MILP surrogate and should
not be interpreted as global optimality with respect to the original
nonlinear profiling model. The dual-annealing solver, in contrast,
provides no global-optimality guarantee.}

Table~\ref{tab:solver-comparison} presents a quantitative comparison of the two solvers implemented in the Reasoner component. Three metrics are reported for each scenario: the average computational time required by the Reasoner, the mean objective value achieved (lower is better for scenarios 1 and 3, higher is better for scenario 2), and the percentage of problem instances where the solver failed to find a feasible dispatch (fail rate). The values in the table are aggregated as means over all tested circuit instances; the full, non-aggregated results for each algorithm and circuit size are illustrated in the figures provided in the previous subsections.

\begin{table}[htbp]
\centering
\begin{tabular}{ll|cc}
\toprule
\textbf{Scenario} & \textbf{Metric} & \textbf{qb-linear} & \textbf{qb-nonlinear} \\
\midrule
\multirow{3}{*}{Scenario 1} 
    & Reasoner Time (s) & 0.021 & 23.913 \\
    & Evaluation    & 249.84 & 215.20 \\
    & Fail Rate (\%)    & 28.57 & 0.00 \\
\midrule
\multirow{3}{*}{Scenario 2} 
    & Reasoner Time (s) & 0.079 & 22.586 \\
    & Evaluation    & 255.46 & 375.35 \\
    & Fail Rate (\%)    & 0.00 & 0.00 \\
\midrule
\multirow{3}{*}{Scenario 3} 
    & Reasoner Time & 0.018 & 15.969 \\
    & Evaluation    & 466.07 & 234.94 \\
    & Fail Rate (\%)    & 14.29 & 0.00 \\
\bottomrule
\end{tabular}
\caption{\textit{Reasoner time, evaluation, and fail rate by scenario and baseline.}}
\label{tab:solver-comparison}
\end{table}

\rev{The fail rate reported in Table~\ref{tab:solver-comparison} refers specifically to the inability of the optimisation solver to return a candidate dispatch. It should therefore be distinguished from post-solve QoS validation. In Scenario~2, for example, the Linear Solver has a solver fail rate of zero because it returns a candidate for every instance; however, those candidates subsequently fail the original reliability constraint when re-evaluated with the Profiler. Hence, solver success does not necessarily imply validated QoS feasibility when an approximate optimisation model is used.}

As expected from their underlying methodologies, the \texttt{qb-linear} solver consistently achieves very low solution times, typically several orders of magnitude faster than the \texttt{qb-nonlinear} dual annealing approach. This speed advantage makes the linear solver highly suitable for interactive or time-constrained applications, and for scenarios where the \qoslanguageacr model is accurately represented by linear approximations.

However, this speed comes at a cost in terms of robustness and fidelity to the original nonlinear profiling model. The fail rate for \texttt{qb-linear} is notably higher in Scenario 1 (28.57\%) and Scenario 3 (14.29\%), reflecting its difficulty in satisfying all QoS constraints when the linearised model diverges from the more complex, nonlinear behaviour of the actual backend profiles. In these cases, the solver is unable to produce any dispatch that meets all requirements. Even when a solution is found, the evaluation metric for \texttt{qb-linear} is sometimes worse than that of \texttt{qb-nonlinear}, due to suboptimal constraint satisfaction or excessive approximation in shot allocation (as observed in Scenario 2 for the fidelity constraint).

In contrast, the \texttt{qb-nonlinear} solver \rev{achieved a zero observed solver-failure rate across the tested scenarios}, with a fail rate of zero in every case. Although it requires significantly more computational time, on the order of tens of seconds per scenario, this investment yields solutions that consistently satisfy all constraints and more more closely reproduce the behaviour represented by the original nonlinear Profiler functions, thanks to the profiling-driven evaluation at each optimisation step. The objective values obtained by \texttt{qb-nonlinear} are superior in all scenarios, reflecting the advantage of a model that dynamically adapts to nonlinearity and context-specific effects.

In summary, the linear solver provides an attractive option when speed is critical and the problem structure is well-behaved, but the dual annealing solver offers greater reliability and solution quality in the presence of complex QoS models or nonlinear backend behaviour. The results in Table~\ref{tab:solver-comparison} clearly illustrate the trade-off between computational efficiency and the fidelity of QoS-aware quantum workload scheduling.

\subsection{Nonlinear Solver Parameter Sensitivity Analysis}




\new{The previous experimental results used the \texttt{qb-nonlinear} solver with 10 independent dual annealing runs and 40 iterations per run. Since these parameters directly affect both the probability of finding high-quality feasible solutions and the computational cost of the solver, we performed a sensitivity analysis to justify this choice and assess the robustness of the solver with respect to parameter selection.}

\new{In particular, we varied the number of independent runs over the set
\(\{1,5,10,15,20,25,30\}\), and the number of iterations per run over the set
\(\{1,5,10,15,\ldots,55,60\}\). For each parameter combination, the solver was executed on all circuit instances in each of the three benchmark scenarios. We then computed, for each scenario, the mean solver execution time and the mean objective function value across all circuits.}

\new{This analysis allows us to characterise the trade-off between runtime and solution quality. It also makes it possible to verify whether the default configuration used in the main experiments, namely 10 independent runs and 40 iterations per run, lies in a stable region of the parameter space where further computational effort provides only marginal improvement.}

\Figure[!ht][width=\textwidth]{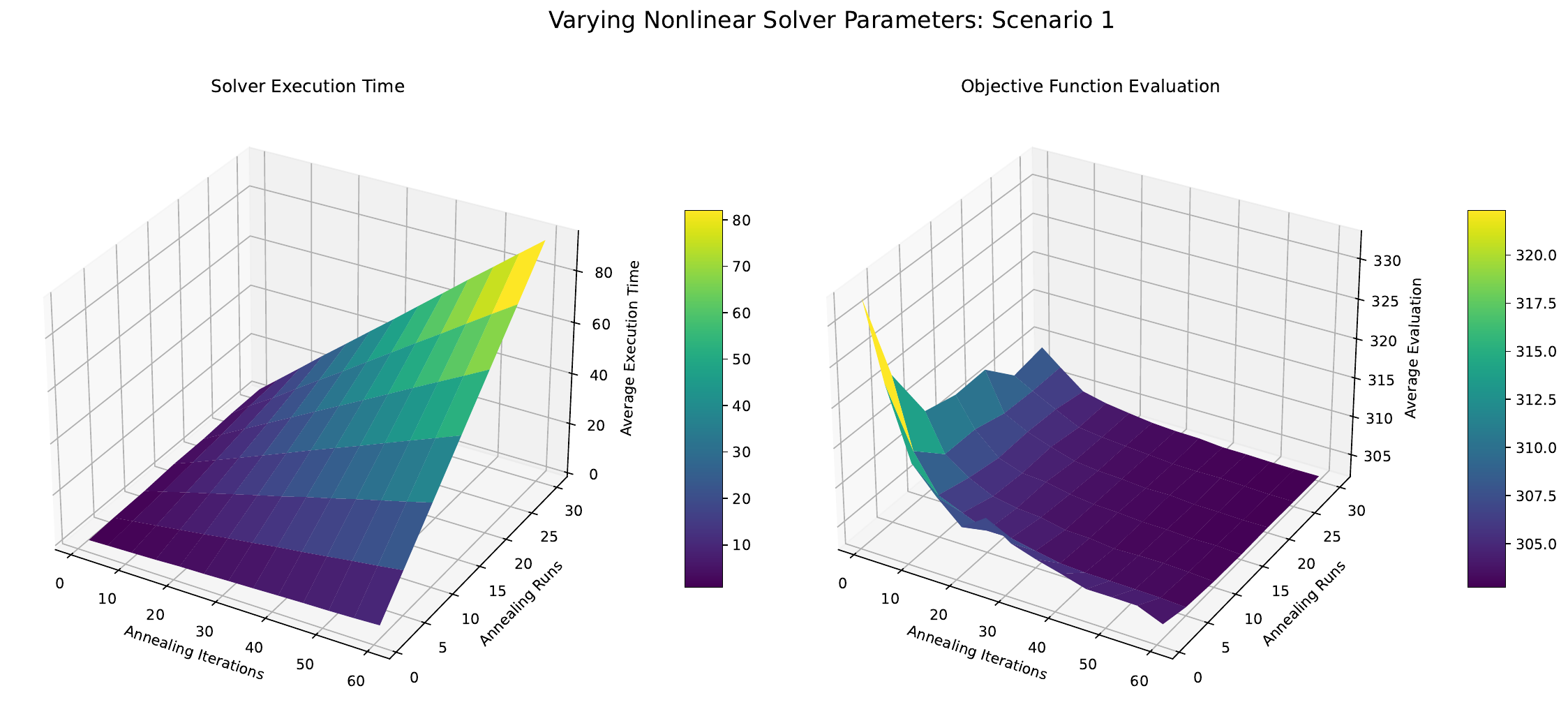}
{Nonlinear solver parameter sweep for Scenario 1. \label{fig:nonlinear-scenario1}} 

Figure~\ref{fig:nonlinear-scenario1} shows mean solver execution time (left panel) and mean objective function value (right panel) as a function of the number of dual annealing iterations and independent runs in Scenario 1. As expected, the solver execution time (left panel) increases almost linearly with the number of independent runs and annealing iterations, reaching values up to approximately 80 seconds for the largest parameter settings. The right panel shows that increasing either parameter initially leads to a clear improvement in the mean objective value (lower scores are better), with diminishing returns beyond a certain point. The surface becomes relatively flat for parameter values greater than about 30 iterations and 10 runs, with the objective stabilising around 303. Notably, this plateau in solution quality is already achieved with execution times between 30 and 40 seconds, indicating that it is not necessary to incur the maximum computational cost \rev{to obtain objective values close to the observed plateau}. This indicates that, for Scenario 1, a practical balance between execution time and solution quality can be achieved at moderate parameter values.

\vspace{1em}

The results for Scenario 2 are depicted in Figure~\ref{fig:nonlinear-scenario2}, in which we observe similar trends. It is important to recall that in this scenario, the objective function is formulated as a maximisation, so higher evaluation values indicate better performance. The solver execution time (left panel) again increases with both the number of runs and iterations, reaching up to 100 seconds at the most intensive parameter values. In contrast to Scenario 1, the mean objective value (right panel) quickly reaches a plateau. Most improvement occurs within 10 to 20 iterations and 5 to 10 runs, after which the objective grows more slowly from 350 to 370, reaching the plateau of about 375 with 30 iterations and 15 runs. At these parameter settings, the execution time is typically in the range of 40 to 60 seconds. This suggests that Scenario 2 is slightly less sensitive to annealing parameter tuning, likely due to broader constraint margins or a simpler optimisation landscape. As before, already achieves objective values close to the observed plateau, enabling considerable computational savings.

\Figure[!ht][width=\textwidth]{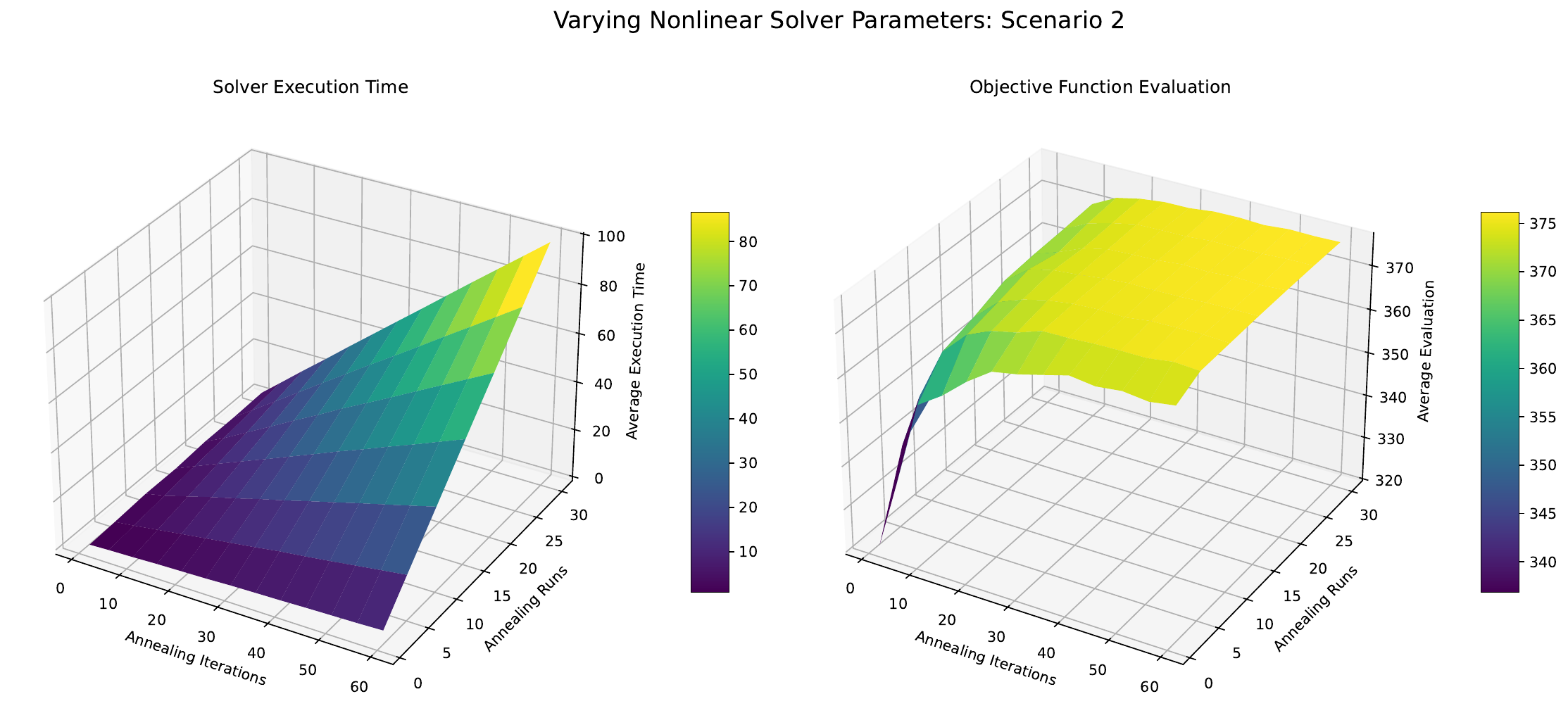}
{Nonlinear solver parameter sweep for Scenario 2. \label{fig:nonlinear-scenario2}} 

\vspace{1em}

Figure~\ref{fig:nonlinear-scenario3} displays the corresponding analysis for Scenario 3. 
Even here, the solver execution time grows with both parameters, reaching about 70 seconds for the highest values. The objective function surface (right panel) shows a noticeable improvement as the number of iterations and runs increases from low to moderate values, with the score dropping sharply up to about 30 iterations and 15 runs. For larger parameter values, the surface rapidly flattens and further improvements become minimal, with the objective stabilising around 265. This plateau is achieved at execution times in the range of 35 to 45 seconds, again \rev{showing that high-quality solutions near the observed plateau can be obtained without maximising computational effort}. \rev{Within the evaluated parameter grid, moderate settings produced stable mean objective values with diminishing returns from additional search effort.}

\Figure[!ht][width=\textwidth]{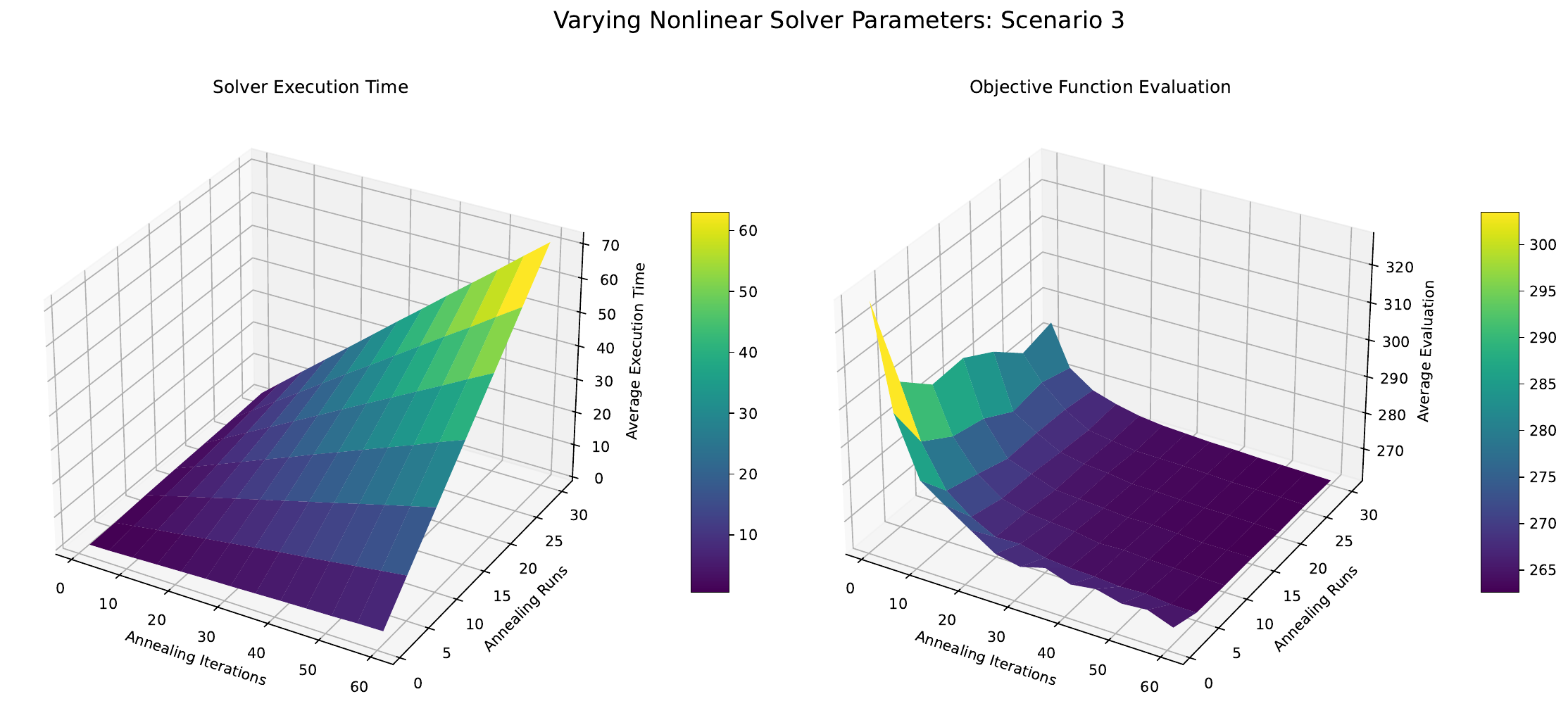}
{Nonlinear solver parameter sweep for Scenario 3. \label{fig:nonlinear-scenario3}} 

\bigskip
Overall, these results illustrate that while increasing the number of dual annealing runs and iterations does increase computational time, substantial improvements in solution quality can be obtained with only moderate parameter values and moderate solver runtimes. For all scenarios, the objective score plateaus once parameters reach intermediate values and solver execution times are still reasonable, allowing users to balance runtime and optimisation quality effectively. In practice, \rev{the Nonlinear Solver showed a zero observed solver-failure rate for the tested instances and configuration} and can be efficiently tuned according to user needs.

\subsection{Discussion}

The experimental results presented above provide a comprehensive assessment of the Quantum Broker and its solvers under a range of realistic and challenging QoS scenarios. Several important observations emerge from this evaluation.

First, the experiments exhibit the crucial value of explicit QoS modelling and automated reasoning in quantum workload dispatch. Across all scenarios, naive baselines that ignore backend heterogeneity or QoS objectives frequently produce infeasible or suboptimal solutions, especially when constraints are tight or conflicting. The ability of the Reasoner to parse flexible \qoslanguageacr models and enforce a wide range of constraints enables it to consistently outperform such baseline strategies, particularly in scenarios with targeted backend assignments or strict requirements on cost, fidelity, and execution time.

Comparing the two solver approaches highlights a clear trade-off between computational efficiency and fidelity to the QoS model used for dispatch planning. The Linear Solver is exceptionally fast and can deliver provably optimal solutions for the linearised version of the \qoslanguageacr problem. This makes it an attractive choice for interactive use or settings where resource constraints and objectives can be well approximated by linear models. However, as the results show, the Linear Solver is more likely to fail in the presence of strong nonlinearity, or to produce candidate allocations that satisfy the linearised constraints but fail the original Profiler-based QoS validation when evaluated with the Profiler. \rev{In particular, the lack of lower bounds on shot assignments in some scenarios, the linearisation of cost and execution-time profiles, and the use of a fixed backend-level approximation for the shot-dependent reliability function are limiting factors of the linear model.}

\rev{The Nonlinear Solver, based on dual annealing, provides a more faithful representation of the nonlinear synthetic Profiler model used in the controlled evaluation. By dynamically evaluating each candidate dispatch with the original Profiler functions throughout the annealing process, it accounts directly for the nonlinear cost, execution-time, and synthetic-reliability relationships represented by that model, rather than relying on the linearised surrogates used by the MILP solver. This distinction concerns fidelity to the profiling model, not to live-QPU operational behaviour. Backend-specific device noise is exercised separately in the end-to-end experiment, whereas the operational QoS quantities used for dispatch planning remain based on controlled profiles.} The parameter sweep experiments confirm that substantial improvements in objective value and constraint satisfaction are achieved for moderate numbers of runs and iterations, with execution times that remain reasonable for most practical scenarios.

\rev{The end-to-end validation complements these planning-level results by evaluating what happens after a dispatch has actually been selected. Across the 18 evaluated nonlinear dispatches, the merged distributions achieve a mean \(F_{\mathrm{E2E}}\) of \(0.907\), demonstrating that the complete shot-wise workflow can preserve high agreement with ideal output statistics even when execution is distributed across heterogeneous noisy backends. The QFT results also highlight that output quality should be assessed at the same aggregation level at which the Quantum Broker returns results to the application: partial executions with very small shot allocations may individually provide poor estimates of a broad probability distribution, while their combined samples still reconstruct a high-quality aggregate distribution.}

\rev{At the same time, the end-to-end experiment clarifies the scope of the conclusions that can be drawn from the current evaluation. The execution results incorporate backend-specific device noise and transpilation constraints, but the operational QoS quantities used by the Reasoner remain based on controlled profiles rather than live provider measurements. Thus, the results provide execution-level evidence for the dispatch and aggregation mechanisms without claiming that the synthetic time, cost, waiting-time, or reliability models reproduce the instantaneous state of production quantum-cloud systems.}

\rev{The present experiments use a fixed pool of five candidate backends and a fixed set of QoS constraints, so the reported runtimes should not be interpreted as a complete empirical scalability study. Nevertheless, the structure of the optimisation problem provides an indication of how its size grows. With \(B\) candidate backends, the basic formulation introduces one shot-allocation variable and one backend-activation variable per backend, while backend-wise QoS constraints and aggregate expressions typically generate additional constraints proportional to the number of backend--constraint combinations. For the nonlinear formulation, the underlying allocation space grows combinatorially: distributing \(S\) shots across \(B\) backends yields \(\binom{S+B-1}{B-1}\) possible non-negative integer allocations. The Nonlinear Solver does not enumerate this space explicitly, but this growth explains why increasing the number of candidate backends and constraints can make the search progressively more difficult. A systematic empirical study varying \(B\) and \(K\) independently is left to future work.}

Overall, these results confirm that the modular Quantum Broker architecture enables a flexible and effective approach to quantum workload scheduling. The choice of solver can be tuned to the requirements of each use case, balancing the need for computational speed with the desire for high-quality, feasible solutions. Moreover, the extensible design of the profiling and evaluation pipeline means that new backends, cost models, and QoS dimensions can be incorporated with minimal effort, supporting both current and future experimental needs. The presented results underscore the value of formal QoS modelling and adaptive reasoning for managing heterogeneous quantum resources.
\section{Known Limitations}
\label{sec:limitations}

To enable a clear understanding of the current capabilities and boundaries of the \textsc{Quantum Broker} framework, we outline here the key limitations identified during design and evaluation. These constraints highlight areas for refinement and represent valuable opportunities for future work.

\begin{enumerate}

    \item \new{\textbf{Synthetic Profiling and Real-Hardware Alignment.} The experimental evaluation relies on noisy simulator backends paired with fixed synthetic QoS profiles. This design enables reproducibility and controlled stress testing, but the profiling formulas for execution time, cost, the synthetic reliability score
are simplified approximations of real hardware behaviour. In particular, execution time is modelled as a function of
circuit depth and shot count, cost is modelled as a nonlinear function of execution time, \(R_{\mathrm{syn}}\) combines a depth-dependent error proxy with a finite-shot uncertainty term. These assumptions capture important qualitative trends, but they do not
fully represent provider-specific compilation overheads, queueing dynamics, calibration drift, device topology,
gate-dependent noise, crosstalk, readout errors, or real pricing policies. Consequently, the reported experiments should be interpreted as evaluating the Broker's reasoning capabilities under
controlled backend heterogeneity, rather than as measuring production performance on live quantum hardware. The
main correction mechanism is empirical profiling: the synthetic parameters can be replaced by values fitted from real
backend observations, and the analytical models can be substituted with more accurate backend-specific predictors.
Future work will include real-hardware validation by collecting execution time, queue delay, cost, and fidelity data
from live devices and comparing the fitted model predictions against held-out quantum circuits.}

    \rev{The end-to-end experiment introduced in this revision strengthens this validation by executing selected dispatches on device-specific noisy simulator backends, merging the returned measurement counts, and evaluating the resulting output distribution against an ideal reference. This allows us to validate the complete dispatch--execution--aggregation workflow under backend-specific noise rather than evaluating only dispatch-time QoS predictions. However, this experiment should not be interpreted as equivalent to live-hardware validation. While the execution outcomes reflect the noise characteristics encoded in the fake backends, operational quantities such as queue delay, execution time, availability, and monetary cost are still obtained from controlled profiles rather than measurements collected from production quantum-cloud services. Empirically measured backend profiles and repeated execution on live QPUs therefore remain important next steps for external validation of the framework.}
    
    \item \textbf{Solver Scalability and Nonlinearity.} The framework supports both a fast mixed-integer linear solver (MILP) and a more robust but slower nonlinear solver (dual annealing). Experimental evidence shows that the linear solver often fails to find feasible solutions when nonlinearity or tight constraints dominate, while the nonlinear solver, though robust, incurs significant computational overhead. This introduces a practical trade-off: the framework cannot guarantee both rapid interactive use and high-fidelity constraint satisfaction under arbitrary workloads, especially as problem size and backend heterogeneity increase. Further, the dual annealing solver does not guarantee global optimality and its performance depends sensitively on parameter selection. \rev{Moreover, the current experimental campaign keeps both the candidate backend pool and the scenario constraint structure fixed. The structural analysis in Section~\ref{sec:experiments} indicates approximately linear growth of the instantiated MILP representation with backend--constraint combinations and combinatorial growth of the underlying shot-allocation space, but an empirical scaling study that systematically varies the number of candidate backends and QoS constraints has not yet been performed.}
    
    \item \textbf{QoS Model Expressiveness and Approximation.} The declarative \qoslanguage enables expressive constraint and objective specification, but its semantics may be approximated by solver models. Linearization can introduce solution infeasibility or hidden constraint violations, particularly for fidelity or cost metrics that scale nonlinearly with shot count or circuit depth. Additionally, the system assumes independence of noise and cost models across backends; correlated errors or unmodeled provider-level effects are not captured. \rev{The post-solve validation step prevents such approximation errors from being silently exposed as QoS-compliant dispatches by re-evaluating candidate allocations with the original profiling model. At present, a candidate that fails this validation is rejected rather than automatically repaired. Developing lightweight repair heuristics that exploit constraint violations and available QoS slack to adjust approximate solutions before resorting to more expensive optimisation is left for future work.}    
    
    \item \rev{\textbf{Shot Independence and Cross-QPU Aggregation.} The Broker relies on independence of repeated shots to make the shot budget distributable, but it does not assume that heterogeneous QPUs implement identical output distributions. Device-specific noise can make each backend sample from a different effective distribution, so naive pooling can be biased with respect to the ideal circuit output even though the individual shots are independent. Accordingly, raw count pooling should be interpreted as estimating the shot-weighted mixture of the participating QPU distributions. Claims of unbiased recovery of a common target require the additional assumption that the QPUs sample the same effective distribution (or have been corrected to that condition). For heterogeneous noisy devices, reliability-aware or bias--variance-aware merge policies, such as those formalized in~\cite{shotwisenoisyarxiv}, are more appropriate when the objective is closeness to the ideal distribution. The end-to-end experiment reported in Section~\ref{sec:end-to-end} quantifies the quality of the current raw-count aggregation strategy for the evaluated workloads, but it does not eliminate this general limitation: different device combinations, stronger heterogeneity, or different output distributions may exhibit larger aggregation bias.}
    
    
    \item \textbf{Extensibility and Maintenance.} Although the architecture is modular and designed for extensibility, integration of new provider APIs, evolving device capabilities, or richer QoS metrics (e.g., energy, geographic location, regulatory compliance) will require ongoing engineering and validation effort. The framework’s future utility depends on active profiling, continued abstraction of provider heterogeneity, and evolution of the \qoslanguageacr and solver infrastructure.
\end{enumerate}

By clearly stating these limitations, our aim is to provide transparency regarding the intended scope of \textsc{Quantum Broker} and motivate further community-driven enhancements that will progressively overcome these constraints in future versions.
\section{Conclusions and Future Work}
\label{sec:concl}

This paper presented the \textit{Quantum Broker}, a modular framework for Quality-of-Service-aware quantum execution management across heterogeneous, cloud-accessible quantum infrastructures. The framework addresses an increasingly important quantum-engineering problem: how to translate user-level execution requirements into concrete dispatch decisions over QPUs that differ in availability, cost, noise behaviour, execution latency, provider interface, and operational constraints. By combining a declarative QoS specification language, backend profiling, constraint-based reasoning, and provider-agnostic orchestration, the Quantum Broker provides an execution-management layer between quantum applications and fragmented NISQ-era quantum cloud resources.

The work operationalises shot-wise distribution as a QoS-aware dispatch strategy. Instead of treating a quantum execution as a fixed submission to a single backend, the framework exploits the statistical independence of measurement shots to allocate portions of a circuit execution across multiple QPUs. This allows execution plans to account explicitly for user-defined objectives and constraints, including cost, execution time, waiting time, fidelity, backend selection, and shot allocation. In this sense, the proposed approach connects a physical property of quantum measurement with a software-level mechanism for adaptive execution management.

\newt{We also introduced \textit{QQSpec}, the declarative Quantum QoS Specification Language used to express execution constraints and optimisation objectives independently of provider-specific APIs. QQSpec provides a compact interface for capturing user intent in terms of cost, fidelity, execution time, waiting time, backend selection, and shot-allocation requirements. By separating QoS specification from backend-specific execution logic, QQSpec supports portability, reproducibility, and explicit reasoning over execution trade-offs.}

The experimental evaluation shows that QoS-aware dispatch can provide clear advantages over naive multi-backend allocation and exhaustive single-backend selection when execution requirements are non-trivial or mutually constraining. The Quantum Broker was able to generate feasible dispatch plans across heterogeneous backend profiles and to expose trade-offs among cost, fidelity, and execution time. These results support the central premise of this paper: execution management should be treated as a first-class concern in quantum software stacks, particularly when applications are deployed over dynamic, multi-provider NISQ infrastructures.

\rev{The complementary end-to-end validation extends these results beyond dispatch planning. For the common subset of workloads for which the Nonlinear Solver produces feasible dispatches in all three scenarios, we executed the selected shot allocations on device-specific noisy simulator backends, merged the returned measurement counts, and compared the resulting aggregate distributions with ideal noise-free references. Across the 18 evaluated executions, the merged distributions achieved a mean squared Bhattacharyya coefficient \(F_{\mathrm{E2E}}\) of \(0.907\), with values ranging from \(0.756\) to \(0.998\). These results provide execution-level evidence that the shot-wise dispatch and aggregation workflow can preserve high output quality under heterogeneous backend-specific noise. They also show that the quality of individual partial executions should not be conflated with the quality of the final result: especially for broad output distributions and small per-backend shot allocations, the merged distribution can provide a substantially better estimate of the target distribution than any individual partial execution.}

The comparison between the linear and nonlinear solvers further clarifies the engineering trade-offs involved in QoS-aware quantum execution. The linear solver offers low planning overhead and is suitable when backend behaviour and QoS objectives can be approximated through linear models. The nonlinear solver, by contrast, is better suited to more complex QoS formulations and heterogeneous backend profiles, at the cost of increased computation time. This trade-off suggests that future quantum execution platforms should support pluggable reasoning mechanisms rather than relying on a single optimisation strategy for all workloads and deployment conditions.

Several limitations remain. The present evaluation uses controlled backend profiles and noisy simulators, which provide reproducibility and controlled heterogeneity but cannot fully capture the temporal variability, calibration drift, queue dynamics, and provider-side constraints of production quantum systems. In addition, the current QoS model focuses on single-circuit shot-wise distribution and does not yet cover richer workload structures such as circuit batches, hybrid quantum--classical workflows, parameter sweeps, or application-level service objectives spanning multiple executions.

\rev{We emphasize that the experiments are not intended to claim numerical performance equivalence with live quantum-cloud execution. The evaluation now covers two complementary levels: dispatch-time QoS reasoning under controlled backend profiles, and execution-level validation of selected dispatches through noisy simulation and aggregation of the returned measurement counts. The latter exercises the complete dispatch--execution--aggregation workflow under device-specific noise, but it does not reproduce all operational characteristics of production quantum-cloud systems. In particular, queue fluctuations, calibration drift, transient availability, provider-side batching, execution latency, and pricing remain represented through controlled profiles rather than live measurements. Real-QPU execution and empirically measured backend profiles therefore constitute important next steps for external validation of both the QoS predictions and the resulting dispatch decisions.}

Future work can therefore focus on integrating real-hardware profiling pipelines and validating the broker under production-grade variability across multiple providers. We also plan to investigate hybrid reasoning techniques, including ML-guided heuristics, adaptive constraint relaxation, and portfolio-based solver selection, to improve scalability while preserving the expressiveness of \qoslanguageacr. \rev{Future evaluation will also consider additional and stronger profile-aware multi-backend heuristic baselines and systematic scalability experiments in which the number of candidate backends and QQSpec constraints is varied independently. This will allow us to quantify both solver-runtime growth and any degradation in feasible-solution quality as the dispatch space becomes larger and more constrained. A further direction is the development of post-solve repair heuristics for approximate solver outputs. Rather than immediately discarding a candidate that fails the exact profiling-model validation, such heuristics could use the detected constraint violations and available slack to redistribute shots and attempt to recover a QoS-feasible dispatch before invoking a more expensive nonlinear optimisation procedure.} Further extensions will target multi-circuit and batch-execution models~\cite{das2019case}, runtime feedback loops, and richer abstractions for application-level QoS. Finally, broader feedback from quantum software engineering practitioners will be used to refine the language, architecture, and tooling so that the Quantum Broker can better support reproducible, portable, and adaptive execution of quantum workloads in cloud-based environments.


\bibliographystyle{IEEEtran}
\bibliography{biblio}

%

\EOD

\end{document}